\documentclass[symmetry,article,accept,moreauthors,pdftex]{Definitions/mdpi} 

\usepackage{booktabs} 
\usepackage{multirow}
\usepackage{soul} 
\usepackage{microtype}
\graphicspath{{Definitions/}}

\usepackage{hyperref}
\makeatletter
\def\UrlAlphabet{%
\do\a\do\b\do\c\do\d\do\e\do\f\do\g\do\h\do\i\do\j%
\do\k\do\l\do\m\do\n\do\o\do\p\do\q\do\r\do\s\do\t%
\do\u\do\v\do\w\do\x\do\y\do\z\do\A\do\B\do\C\do\D%
\do\E\do\F\do\G\do\H\do\I\do\J\do\K\do\L\do\M\do\N%
\do\O\do\P\do\Q\do\R\do\S\do\T\do\U\do\V\do\W\do\X%
\do\Y\do\Z}
\def\UrlDigits{\do\1\do\2\do\3\do\4\do\5\do\6\do\7\do\8\do\9\do\0}
\g@addto@macro{\UrlBreaks}{\UrlOrds}
\g@addto@macro{\UrlBreaks}{\UrlAlphabet}
\g@addto@macro{\UrlBreaks}{\UrlDigits}
\makeatother

\firstpage{1} 
\pubvolume{xx}
\issuenum{1}
\articlenumber{5}
\pubyear{2020}
\copyrightyear{2020}
\history{Received: 21 August 2020; Accepted: 10 September 2020; Published: date}
\updates{yes} 

\usepackage{booktabs} 
\usepackage{multirow}
\usepackage{soul} 
\usepackage{microtype}
\graphicspath{{Definitions/}}

\Title{Very Long Baseline Interferometry Observations of the Proposed Radio Counterpart of an EGRET Source}

\Author{Patrik Milán Veres $^{1,}$*\orcidC{}, Krisztina Éva Gabányi $^{1,2,3,}$*\orcidA{}, and~Sándor Frey $^{2,4}$\orcidB{}}

\AuthorNames{Patrik Milán Veres, Krisztina Éva Gabányi and~Sándor Frey}

\address{%
$^{1}$ \quad Department of Astronomy, Eötvös Loránd University, Pázmány Péter sétány 1/A, \mbox{H-1117 Budapest, Hungary}\\
$^{2}$ \quad Konkoly Observatory, Research Centre for Astronomy and~Earth Sciences, Konkoly Thege Miklós út 15-17,
H-1121 Budapest, Hungary; frey.sandor@csfk.mta.hu\\
$^{3}$ \quad Extragalactic Astrophysics Research Group, Eötvös Loránd University, Pázmány Péter sétány 1/A, \mbox{H-1117 Budapest, Hungary}\\
$^{4}$ \quad Institute of Physics, ELTE Eötvös Loránd University, Pázmány Péter sétány 1/A, H-1117 Budapest, Hungary}

\corres{Correspondence: patrik1885@student.elte.hu (P.M.V.); gabanyi@konkoly.hu (K.É.G)} 

\abstract{We present high-resolution radio interferometric imaging observations of the radio source NVSS\,J182659$+$343113 (hereafter J1826$+$3431), the~proposed radio counterpart of the $\gamma$-ray source, 3EG\,J1824$+$3441 detected by the Energetic Gamma Ray Experiment Telescope (EGRET) on board the {\it Compton Gamma Ray Observatory} satellite. We~analyzed eight epochs of archival multi-frequency very long baseline interferometry data. We~imaged the asymmetric core--jet structure of the source, and~detected apparent superluminal motion in the jet. At~the highest observing frequency, $15.3$\,GHz, the~core shows high brightness temperature indicating Doppler boosting. Additionally, the~radio features undergo substantial flux density variability. These findings strengthen the previous claim of the association of the blazar J1826$+$3431 with the possible $\gamma$-ray source, 3EG\,J1824$+$3441.} 


\begin{document}


\section{Introduction}
Blazars are radio-loud active galactic nuclei (AGN) seen at a very small angle to the jet direction. The~relativistic speed of the emitting plasma in the jet and~the small angle to the line of sight give rise to various relativistic effects such as apparent superluminal motion and~relativistic Doppler boosting of the radiation. Due to the latter, the~emission of the approaching jet is enhanced while that of the receding jet (also called counterjet) is diminished significantly. The~jet-to-counterjet flux density ratio can be as high as $1000$~\cite{unified_model}. As a consequence, only the approaching one of the originally symmetrically launched jets can be observed, thus blazars appear asymmetric in that regard. The~jet emission is explained as synchrotron radiation from electrons and/or positrons moving with relativistic speeds in magnetic fields. The~jet synchrotron radiation is responsible for the lower-energy bump seen in the broad-band spectral energy distribution (SED) of blazars, peaking between millimetre-wavelength and~X-ray regimes~\cite{Giommi2012}.

Blazars constitute the most populous group of extragalactic $\gamma$-ray emitter sources according to the observations of the {\it Fermi} Large Area Telescope (LAT; e.g.,~\cite{4FGL_cat}). The~$\gamma$-ray emission, responsible for the higher-energy bump in the blazar SEDs, is usually attributed to inverse-Compton process in leptonic models (e.g.,~\cite{Giommi2012} and~references therein) or to  synchrotron radiation from heavy charged particles, e.g.,~protons (\cite{hadronic} and~references therein) in hadronic models. The~seed photons of the inverse-Compton radiation can originate from within or external to the jet. In the former case, they are the low-energy synchrotron photons in the jet (i.e., synchrotron self-Compton process~\cite{ssc}), or in the latter case, the~seed photons can come from the accretion disk, the~broad line region, from the dusty torus or from the cosmic microwave background radiation.

Prior to {\it Fermi}, the~EGRET instrument on board the {\it Compton Gamma Ray Observatory} (CGRO) satellite created the first all-sky $\gamma$-ray map above $100$\,MeV. The~third EGRET catalogue (3EG,~\cite{3EG}) contains $271$ sources, $94$ of them showed possible or probable associations with blazars~\cite{Thompson2008}. Follow-up studies (e.g.,~\cite{new_id} and~references therein) significantly increased the number of identified northern hemisphere EGRET sources at high Galactic latitudes ($|b|>10^\circ$). More than half of the $116$ northern EGRET sources, $66$, are found to have plausible blazar-like counterparts~\cite{new_id}.

One of them, 3EG\,J1824$+$3441, is associated with a flat-spectrum radio quasar at a redshift \mbox{$z=1.81$.} This radio source was detected in the NRAO VLA Sky Survey (NVSS,~\cite{nvss}). J1826$+$3431 has a 1.4-GHz flux density of $(470 \pm 14)$\,mJy which is the highest value in the EGRET positional error box \mbox{($\sim1^{\circ}$)~\cite{EGRET_pos}.} Since the source belongs to the Very Long Baseline Array (VLBA) Calibrator Survey~\cite{vcs2} sample, high-resolution radio interferometric observations of the source exist, which can be used to strengthen or falsify its inferred blazar nature and~in turn its association with the EGRET $\gamma$-ray source. The accurate radio position (\url{http://hpiers.obspm.fr/icrs-pc/newwww/icrf/icrf3sx.txt}) 
of J1826+3431 in the current 3rd realization of the International Celestial Reference Frame~\cite{icrf3} is right ascension $18^\mathrm{h} 26^\mathrm{min} 59.9828210 \pm 0.0000091^\mathrm{s}$ and~declination $34^\circ 31^{\prime} 14.119974 \pm 0.00019^{\prime\prime}$.

Throughout this paper, we assume a flat $\Lambda$CDM cosmological model with $H_0=70\mathrm{\,km\,s^{-1}\,Mpc^{-1}}$, $\Omega_\Lambda=0.73$, and~$\Omega_\mathrm{m=}0.27$. In this model, the~luminosity distance of the source at $z=1.81$ is 14\,103.7\,Mpc, and~1 milli-arcsecond (mas) angular size corresponds to 8.66\,pc projected linear size~\cite{cosmo-calc}.

 
\section{Archival Radio Data}
\unskip
\subsection{VLBI Data}

In the very long baseline interferometry (VLBI) image database of the astrogeo.org website \mbox{(\url{http://astrogeo.org/} maintained by L. Petrov),} there are four epochs of dual-band \mbox{$2.3$- and~$8.3/8.7$-GHz} observations and~three epochs of $15.3$ GHz measurements available of J1826$+$3431 between $1996$ and~$2018$. In each of these observations, all ten antennas of the VLBA participated. Further details of the observations are given in Table~\ref{tab:obs}.

\begin{table}[H]
\caption{Summary of archival VLBI observations of J1826$+$3431. Asterisk (*) denotes the observation conducted by the EVN. All other observations were done by the VLBA. \label{tab:obs}}
\centering
\begin{tabular}{ccccc}
\toprule
\textbf{\multirow{2}{*}{Epoch}}	& \textbf{\multirow{2}{*}{Project ID}} & \textbf{Frequency}	& \textbf{Bandwidth} & \textbf{On-Source Integration Time}\\
 & & \textbf{(GHz)} & \textbf{(MHz)} & \textbf{(min)} \\
\midrule
$1996.370$	& BB023	& $2.3$			& $32$ & $50.3$\\
$1996.370$	& BB023	& $8.3$			& $32$ & $50.3$ \\
$2003.655$	& BU026	& $15.3$			&$16$ & $325.3$ \\
$2004.616$	& BU026	& $15.3$			& $16$ & $325.3$ \\
$2005.427$	& BU026	& $15.3$			& $16$ & $328.1$\\
$2014.145$ & EF025 * & $1.7$ & $128$ & $15.0$ \\ 
$2015.064$	& BG219	& $2.3$			& $128$ & $1.8$ \\
$2015.064$	&	BG219 & $8.7$			& 320 & $1.8$ \\
$2017.227$	& UF001	& $2.3$			& 96 & $2.2$ \\
$2017.227$	& UF001	& $8.7$			& 384 & $2.2$ \\
$2018.437$	& UG002	& $2.3$			& 96 & $2.6$ \\
$2018.437$	& UG002	& $8.7$			& 384 & $2.6$ \\
\bottomrule 
\end{tabular}
\end{table}

\newpage
Additionally, the~source was observed with the European VLBI Network (EVN) at $1.7$\,GHz on 21 February 2014 (project code: EF025, PI: S. Frey). This observation was conducted in phase-referencing mode~\cite{phase-ref}, and~J1826$+$3431 was used as the phase-reference calibrator source for one of the faint targets in that project~\cite{ef025}. Eight antennas of the EVN participated in the measurements: Effelsberg in Germany, the~Jodrell Bank Lovell Telescope in the United Kingdom, Medicina and~Noto in Italy, Onsala in Sweden, Toru\'n in Poland, the~Westerbork Synthesis Radio Telescope in the Netherlands, and~Sheshan in China. Further details of the observation are given in Table~\ref{tab:obs} and~in~\cite{ef025}.

\subsection{VLA Data}

The public data archive (\url{https://archive.nrao.edu/}) of the U.S. National Radio Astronomy Observatory (NRAO) lists various data sets which contain VLA observations of J1826+3431. To~obtain supplementary information on the arcsec-scale extended radio structure, as well as the possible variability of the source, we selected two data sets obtained at 4.86\,GHz for analysis. The~experiment AS291 (PI: W.C. Saslaw) was observed at the epoch 1987.493 in the most extended A configuration of the array that provides the highest angular resolution, in our case, $\sim$0.5\,arcsec. The~second experiment from 2009.388 (project code: CALSUR) used the less extended VLA B configuration, providing resolution of a few arcsecs. The~integration times spent on J1826+3431 were 219\,s and~43\,s in 1987 and~2009, respectively. Both experiments used 100\,MHz bandwidth. 

\section{Data Analysis}
\unskip
\subsection{VLBA Data}
\label{vlba data}

Calibrated visibility files were downloaded from the astrogeo.org website, and their hybrid mapping procedure was performed using the Caltech {\sc Difmap}~\cite{difmap} software package. Asterisks (*) denote those epochs in Tables \ref{tabLk}--\ref{tabUk} where we time-averaged the calibrated data into $10$-s blocks. In a process of hybrid mapping, the~{\sc clean} algorithm~\cite{clean} was used and~phase-only self-calibration was done. At~the end of the procedure, steps of amplitude and~phase self-calibration were performed, gradually for shorter and~shorter solution time intervals down to a few minutes. Images (Figure~\ref{fig:maps}) were created using natural weighting, with the visibility errors raised to the power $-1$.

\subsection{EVN Data}
\label{evn data}

The EVN data calibration was done in the usual manner (e.g.,~\cite{aips_datared}) using the NRAO Astronomical Image Processing System ({\sc AIPS},~\cite{aips}). The~detailed description is given in~\cite{ef025}. The~calibrated data set was imaged in {\sc Difmap}~\cite{difmap} using the hybrid mapping technique, which involved several steps of {\sc clean} iterations interleaved by phase-only self-calibrations. When the signal-to-noise ratio in the residual image could not be improved any further, amplitude and~phase self-calibration were performed. First, it was done for the whole observation; afterwards, the corrections were computed for a subset of antennas and~with gradually decreasing solution time intervals down to $10$~min. The~image obtained is displayed in Figure~\ref{fig:maps}a.

\subsection{VLA Data}
\label{vla data}

\begin{figure}[H]
\begin{minipage}[t]{0.48\textwidth}
\centering
\includegraphics[width=.95\textwidth, bb=160 85 565 490, clip=,]{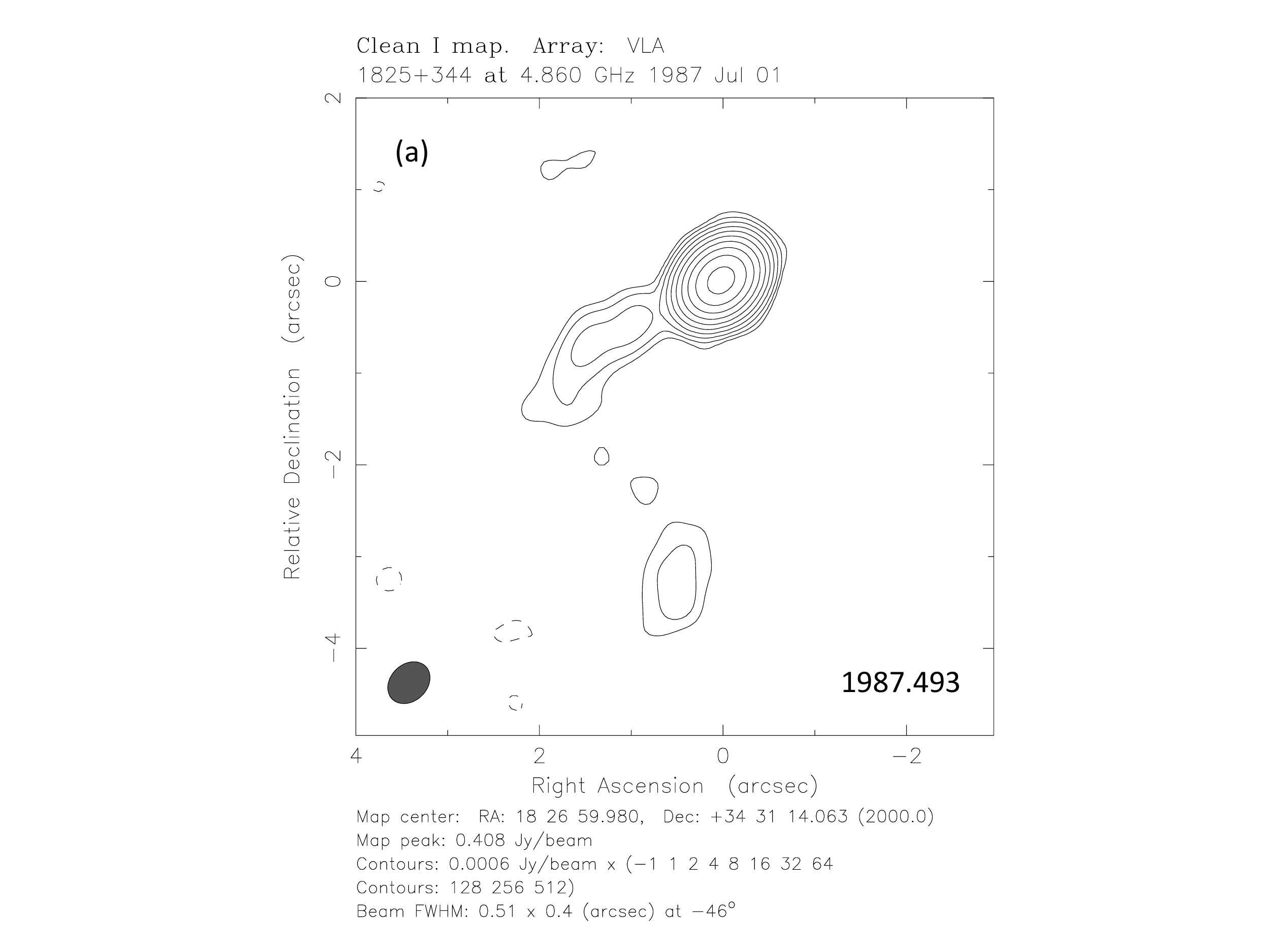}
\end{minipage}
\hfill
\begin{minipage}[t]{0.48\textwidth}
\centering
\includegraphics[width=.95\textwidth, bb=160 85 565 490, clip=,]{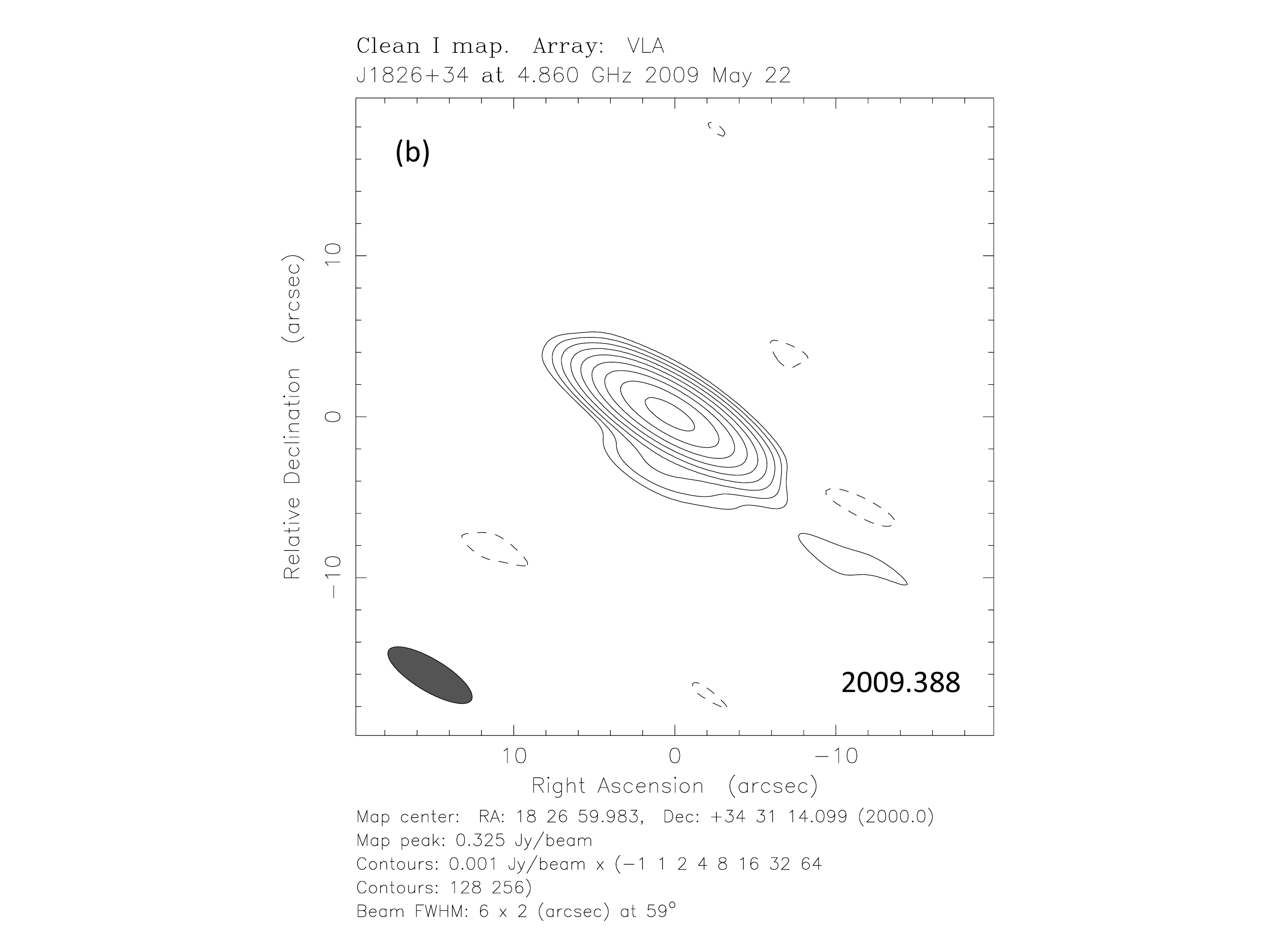}
\end{minipage}
\caption{\label{fig:vla-maps}Naturally-weighted 4.86-GHz VLA images of J1826$+$3431 taken with (\textbf{a}) the A configuration and~(\textbf{b}) the B configuration of the array at the epochs 1987.493 and~2009.388, respectively.  The~lowest contours are (\textbf{a}) $\pm 0.6$\,mJy\,beam$^{-1}$ and~(\textbf{b}) $\pm 1$\,mJy\,beam$^{-1}$. The~peak brightnesses are \mbox{(\textbf{a}) $408$\,mJy\,beam$^{-1}$} and~(\textbf{b}) $325$\,mJy\,beam$^{-1}$. The~restoring beams are (\textbf{a}) $0.51^{\prime\prime} \times 0.4^{\prime\prime}$ at PA=$-46^\circ$ and~(\textbf{b}) $6^{\prime\prime} \times 2^{\prime\prime}$ at PA=$59^\circ$. In both images, the~elliptical Gaussian restoring beam (FWHM) is shown at the lower left corner and~the positive contours increase by a factor of 2.}
\end{figure}

We performed standard amplitude and~phase calibration in {\sc AIPS}. The~VLA flux density calibrator to set the amplitude scale was 3C\,48 in both experiments. The~calibrated visibility data were exported from {\sc AIPS} to {\sc Difmap} for hybrid mapping. The~resulting {\sc clean} images are shown in Figure~\ref{fig:vla-maps}.

\newpage
\subsection{Model Fitting}
\label{model fitting}

To describe the brightness distribution for a quantitative analysis of the source structure, we~applied the technique of model fitting directly to the calibrated interferometric visibility data~\cite{modfit}, for both the VLBI and~VLA measurements. We~used {\sc Difmap} to represent the source brightness distribution with a set of circular Gaussian model components.

\begin{figure}[H]
\begin{minipage}[t]{0.48\textwidth}
\centering
\includegraphics[width=\textwidth, clip=, bb=160 75 545 485]{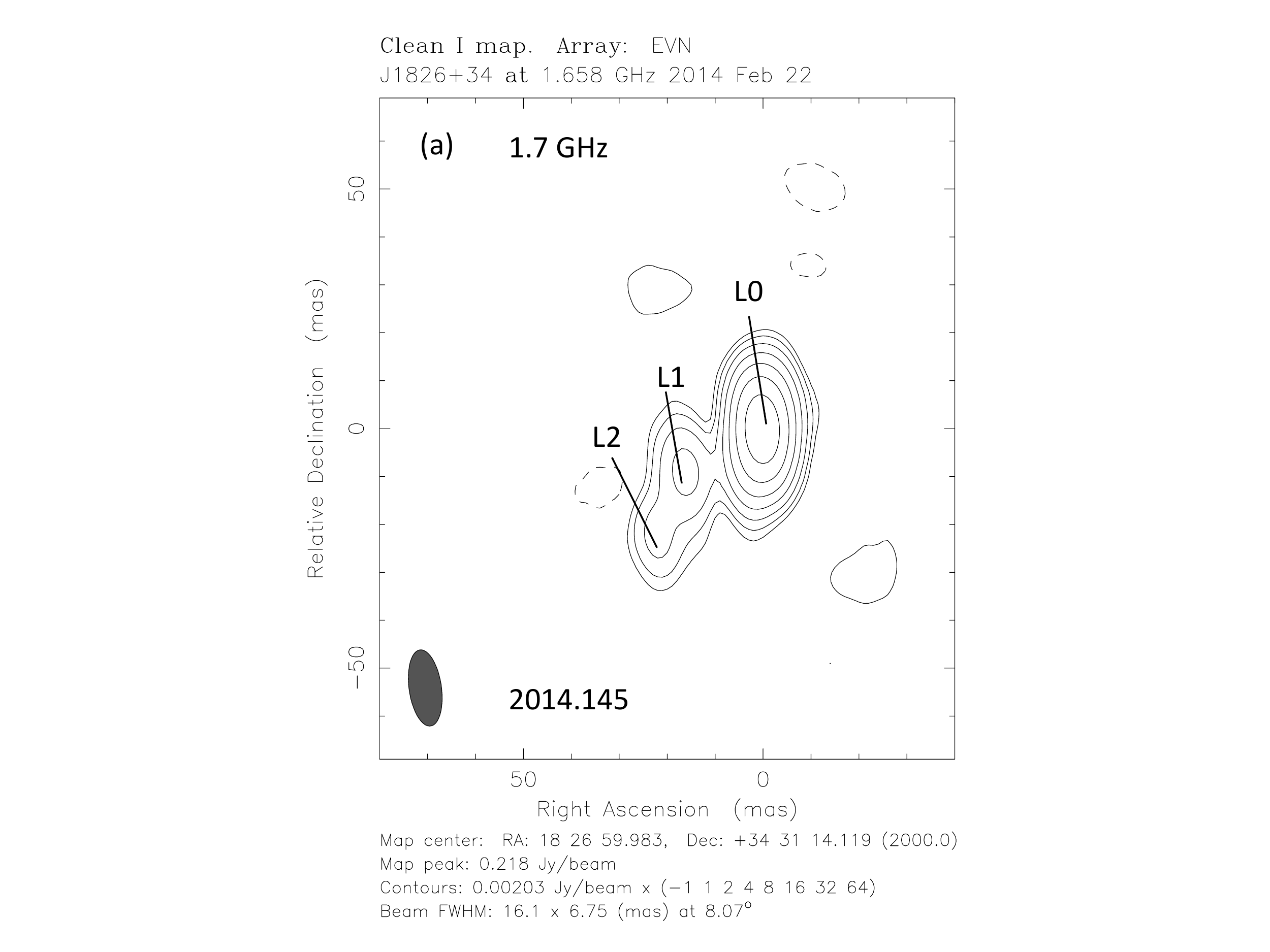}
\end{minipage}
\begin{minipage}[t]{0.51\textwidth}
\centering
\includegraphics[width=\textwidth, clip=, bb=160 70 565 485]{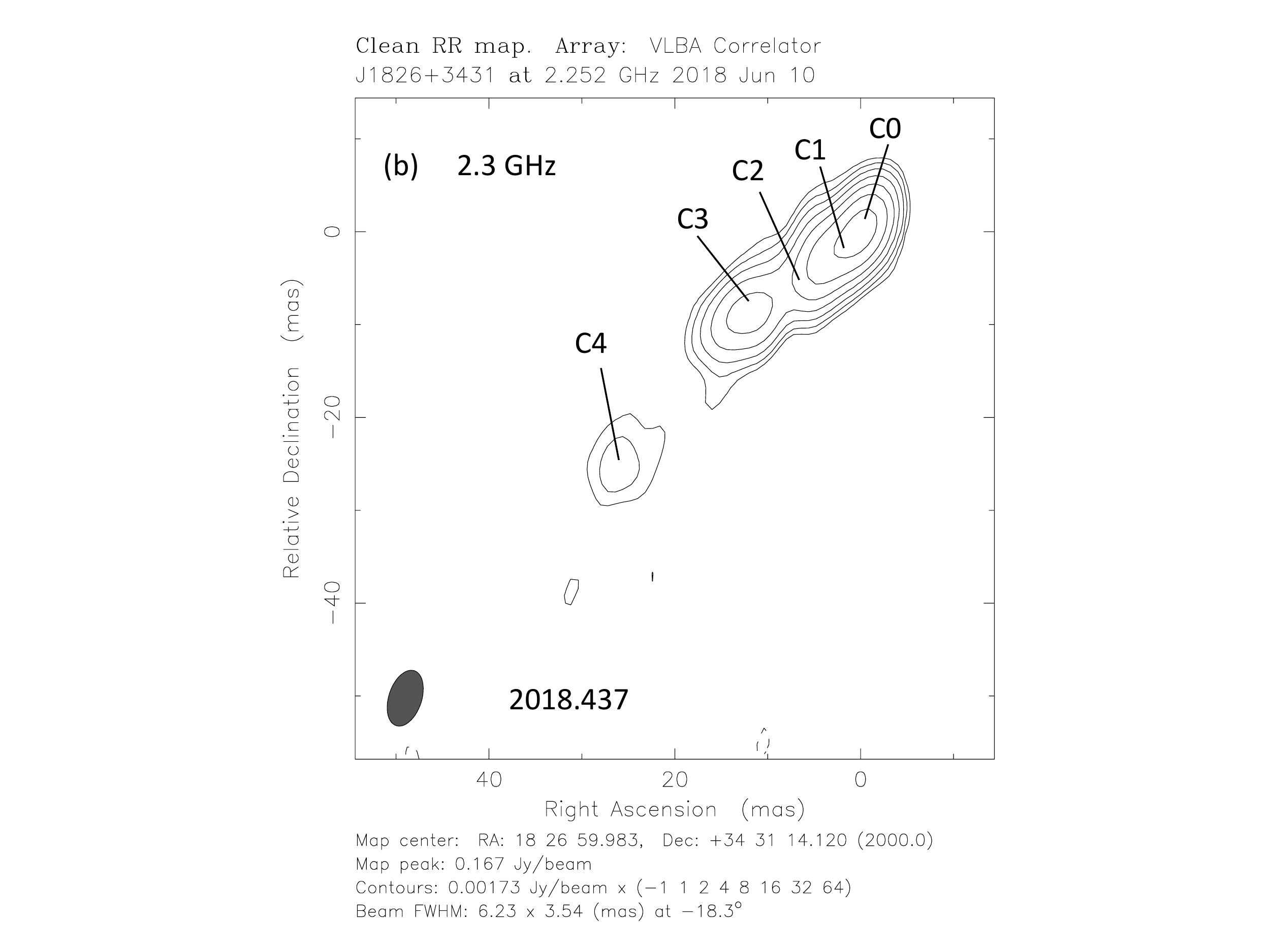}
\end{minipage}
\hfill
\begin{minipage}[t]{0.52\textwidth}
\centering
\includegraphics[width=\textwidth, clip=, bb=120 70 610 485]{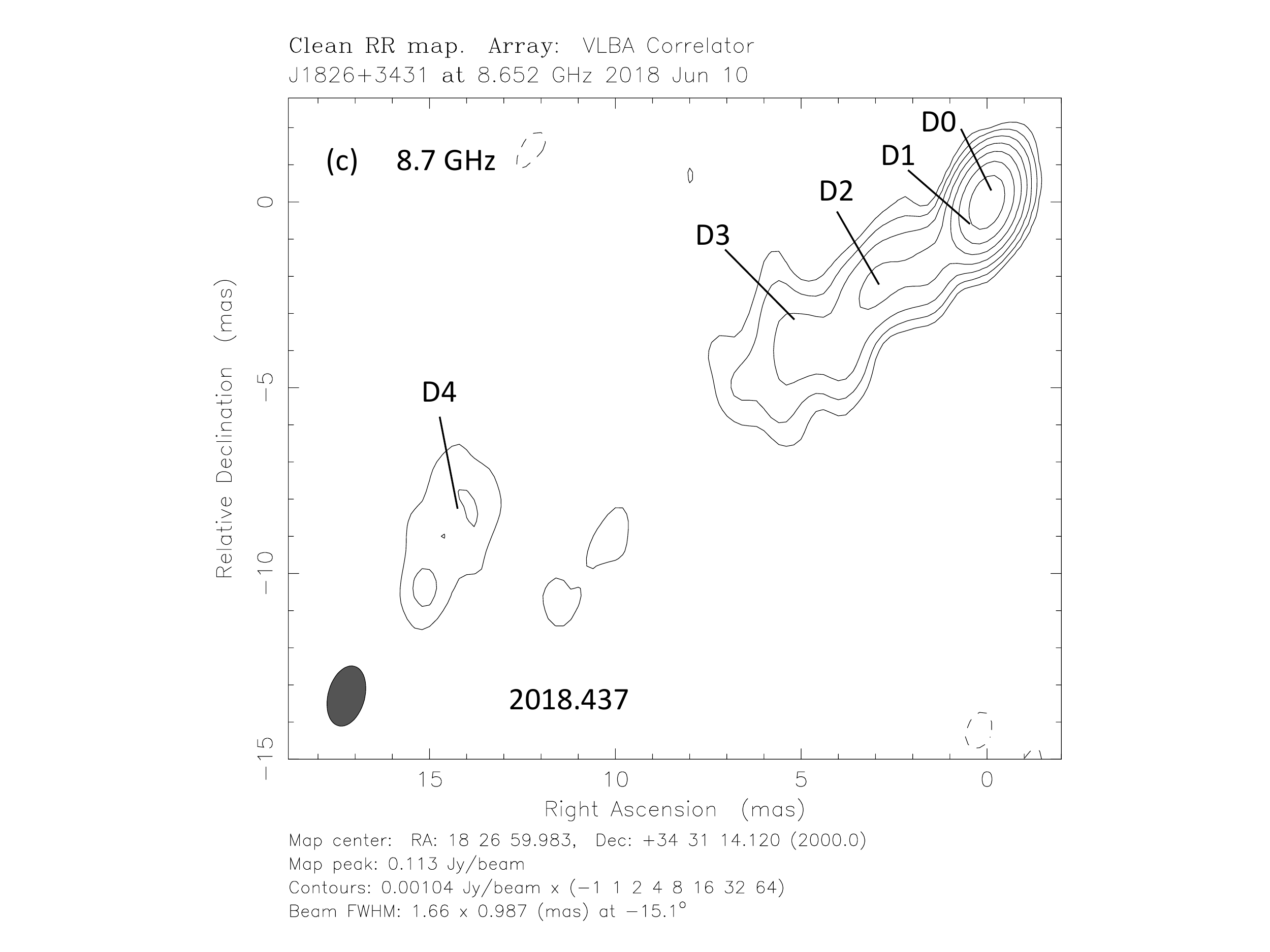}
\end{minipage}
\begin{minipage}[t]{0.48\textwidth}
\centering
\includegraphics[width=\textwidth, clip=, bb=140 70 570 490]{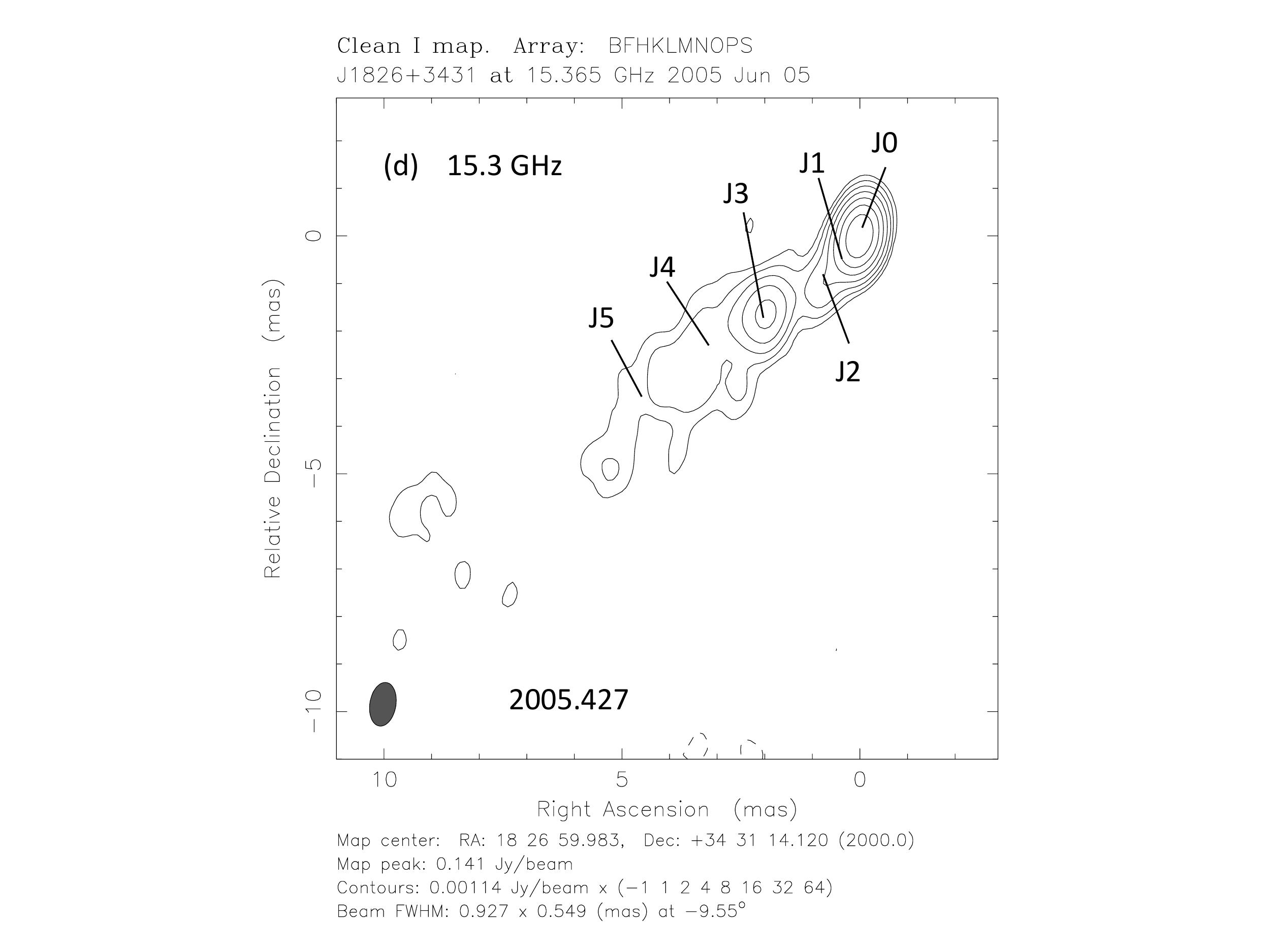}
\end{minipage}
\caption{\label{fig:maps}Naturally-weighted VLBI images of J1826$+$3431 taken at (\textbf{a}) $1.7$\,GHz; (\textbf{b}) $2.3$\,GHz; (\textbf{c}) $8.7$\,GHz; and~(\textbf{d}) $15.3$\,GHz. 
In each image, the~elliptical Gaussian restoring beam (FWHM) is shown in the lower left corner. The~positions of the circular Gaussian model components describing the brightness distributions, observational epochs, and~frequencies are indicated in each image. Further details of the images are given in Table~\ref{tab:maps_details}.}
\end{figure}  
\unskip
\unskip
\begin{table}[H]
\caption{\label{tab:maps_details} Details of the contour maps shown in Figure~\ref{fig:maps}. The~restoring beam major axis position angle (PA) is measured from north through east. Further positive contour levels in Figure~\ref{fig:maps} increase by a factor of 2.}
\centering
\tablesize{\footnotesize}
\begin{tabular}{cccccccc}
\toprule
\vspace{1pt}

 & & \textbf{Peak} & \textbf{Image} & \textbf{Lowest} & \multicolumn{3}{c}{\textbf{Restoring Beam}} \\ \cline{6-8}
\vspace{1pt}
\textbf{Epoch}	& \textbf{Frequency}	& \textbf{Brightness} & \textbf{Noise Level} & \textbf{Contour} & \textbf{Major Axis} & \textbf{Minor Axis} & \textbf{PA}\\ 

 & \textbf{(GHz)} & \textbf{(mJy\,beam}\boldmath{$^{-1}$)} & \textbf{(mJy\,beam}\boldmath{$^{-1}$)} & \textbf{(mJy\,b
eam}\boldmath{$^{-1}$)} & \textbf{(mas)} & \textbf{(mas)} & \textbf{(\boldmath{$^\circ$})}\\
\midrule
$2014.145$		& $1.7$			  & $218$ & $0.4$ & $\pm2.0$ & $16.10$ & $6.75$ & $8.1$ \\
$2018.437$		  & $2.3$			& $167$ & $0.4$ & $\pm1.7$ & $6.23$ & $3.54$ & $-18.3$ \\
$2018.437$		  & $8.7$            & $113$ & $0.2$ & $\pm1.0$ & $1.66$ & $0.99$ & $-15.1$ \\
$2005.427$		& $15.3$			& $141$ & $0.4$ & $\pm1.1$ & $0.93$ & $0.55$ & $-9.6$ \\
\bottomrule
\end{tabular}
\end{table}

\section{Results}

Based on VLBI data, the~inner structure of J1826$+$3431 shows a southeast oriented core--jet morphology consistently at all frequencies and~epochs (see Figure~\ref{fig:maps} for selected example maps). The~core and~typically 4--5 jet features were needed to adequately model the source with circular Gaussian components at $2.3$, $8.3/8.7$, and~$15.3$\,GHz. At~the lowest frequency, 1.7\,GHz, only two model components were necessary to describe the jet brightness distribution. During the fitting procedure, all of the parameters, i.e., positions, flux densities, and~full width at half-maximum (FWHM) sizes of the components were allowed to vary, except for one epoch. At~$2005.427$, the~fitting procedure was not able to constrain the FWHM size of the core component; therefore it was fixed at the smallest angular size resolvable with the interferometer, $0.04$\,mas. To~calculate this value, we used the formula given by~\cite{smallest_size}. Thus, we can only give an upper limit of the size of the core component at that epoch. The~parameters of the best-fit models are summarized in Tables~\ref{tabLk}--\ref{tabUk}; the model component positions are also denoted in the images in Figure~\ref{fig:maps}.

When deriving the uncertainties of the fitted parameters of the VLBI observations, we assumed $10\%$ flux density calibration uncertainties following~\cite{hibak}. The~positional uncertainties were calculated conservatively following~\cite{poshiba}. We~quadratically added the uncertainties determined using the relations given by~\cite{hiba_kepletek} to the $10\%$ of the restoring beam size in the cases of the core components, and~to the $20\%$ of the restoring beam size when deriving the positional uncertainty of less bright jet components. To~calculate the uncertainties of the FWHM sizes, we used the formula of~\cite{hiba_kepletek}.

The arcsec-scale radio structure of J1826+3431 revealed by our 4.86-GHz VLA images (Figure~\ref{fig:vla-maps}) indicates a continuing jet towards the southeast, in approximately the same position angle as the mas-scale VLBI jet (Figure~\ref{fig:maps}). The~jet can be traced out to $\sim$2.5$^{\prime\prime}$ (corresponding to $\sim$20, kpc projected distance from the core) in the higher-resolution A-configuration image. There is an indication of a sharp right-angle bend towards south, as supported by the lower-resolution B-configuration VLA image where the radio emission seems extended in that direction (Figure~\ref{fig:vla-maps}). The~sum of the flux densities in the fitted Gaussian model components ($428 \pm 21$\,mJy at 1987.493 and~$333 \pm 17$\,mJy at 2009.388) provide evidence for significant variability.

\section{Discussion}
\unskip
\subsection{Component Proper Motions in the Mas-Scale Jet of J1826$+$3431}

To identify the jet components throughout the full time span of the VLBI observations, we~considered their positions, FWHM sizes, and~flux densities. The~components that are identified across the different epochs are denoted by the same identifiers in Tables~\ref{tabSk}--\ref{tabUk}. The~core component is numbered with 0 at each epoch and~frequency; thus, it is denoted as L0, C0, D0, and~J0 at $1.7$, $2.3$, $8.3/8.7$, and~$15.3$\,GHz, respectively. 

For further analysis, we assumed the core to be stationary and~we derived the positions of the fitted jet components relative to it (Tables~\ref{tabLk}--\ref{tabUk}). The~jet components are assumed to move away from the core at constant speeds along a straight trajectory. Thus, we fitted their core separations with linear functions. The~derived angular proper motions were converted to apparent speeds in units of the speed of light, $c$, using the relation~\cite{unified_model}
\begin{equation}
\beta_{\textrm{app}} = \frac{\mu \, d_{\textrm{L}}}{c (1+z)}\,
\end{equation}
where $\mu$ is the angular proper motion of the component in rad\,s$^{-1}$, $d_{\textrm{L}}$ is the luminosity distance in m, $c$ is the speed of light in m\,s$^{-1}$, and~$z$ is the redshift of the source.

\begin{figure}[H]%
\begin{minipage}[t]{0.49\textwidth}
\centering
    \includegraphics[width=\textwidth, bb=10 40 710 530, clip=]{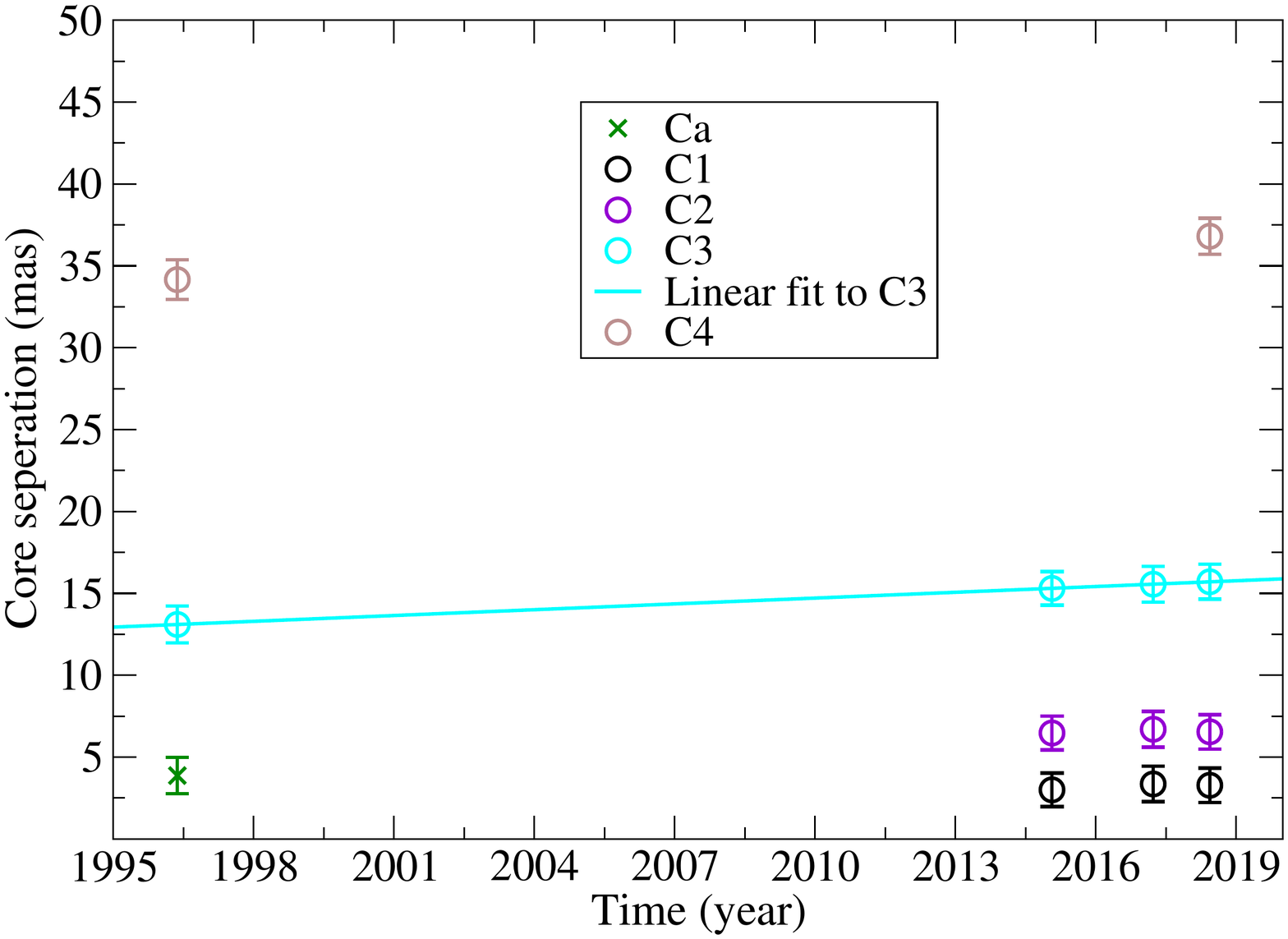}
\end{minipage}
\begin{minipage}[t]{0.49\textwidth}
    \centering
    \includegraphics[width=\textwidth, bb=10 40 710 530, clip=]{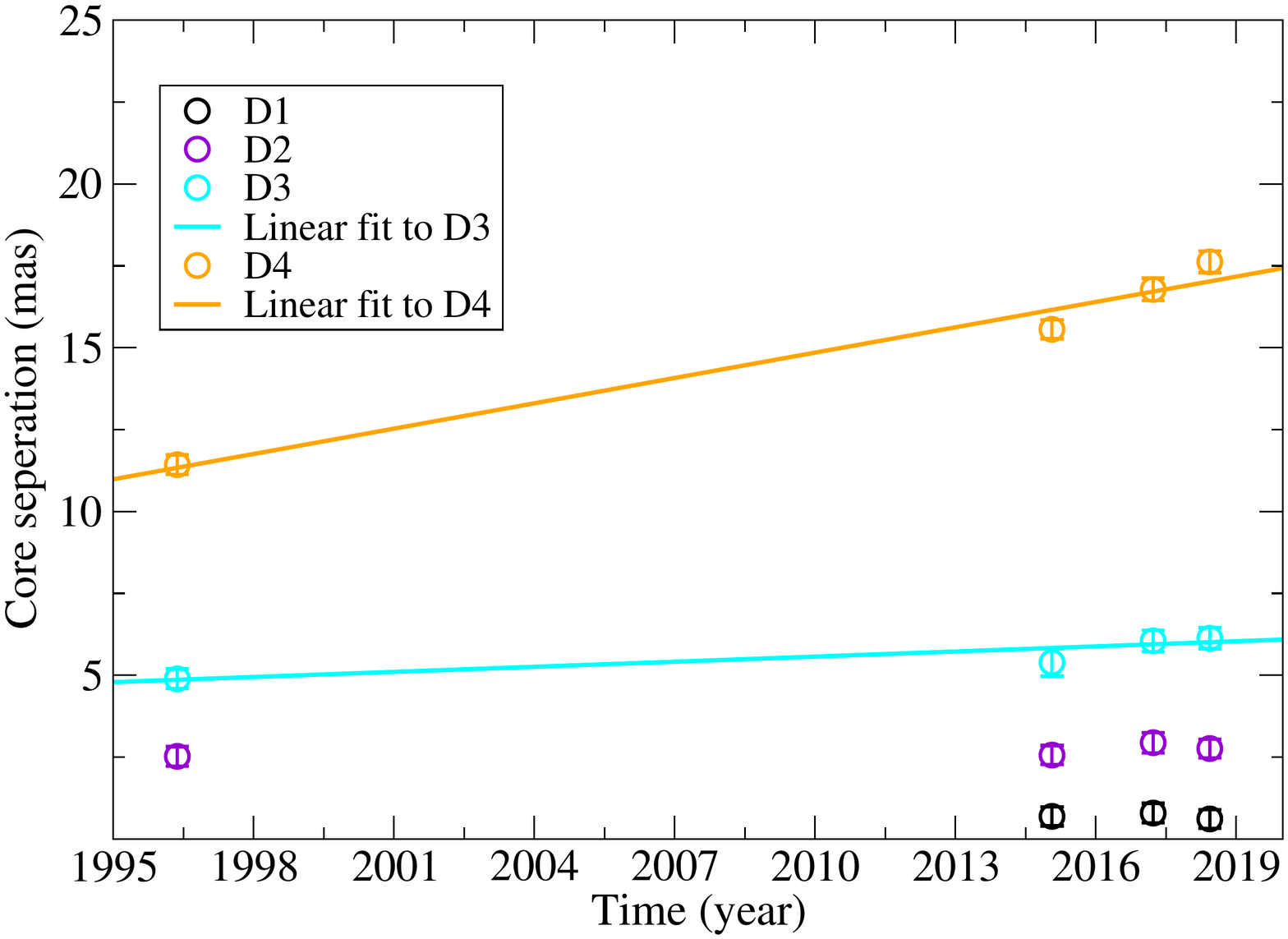}
\end{minipage}
    \caption{\label{fig:SXtav}Radial separations of the components from the core as a function of time; solid lines denote the linear fit to the data. {\it Left panel:} For the observations conducted at $2.3$\,GHz {\it Right:} For the observations conducted at $8.3/8.7$\,GHz. The size of the error bars is roughly the same as the size of the symbols.} 
    \end{figure}

At $2.3$\,GHz, at most five components were needed to describe the jet structure. The~inner $\leq$10\,mas region of the jet could not be resolved into two distinct components in $1996.370$, unlike at later epochs (components C1 and~C2). Therefore, the~single fitted inner jet component, Ca, cannot be unambiguously identified with either of the components detected later. Thus, the~motion of C1 and~C2 can only be tracked at the relatively closely-spaced observing epochs of the 2010s. Because of the short time baseline, these two components show no significant change in their core separation during the available observing epochs. On the other hand, component C3 could be unambiguously identified through all epochs. It shows an angular proper motion of $\mu = 0.12 \pm 0.07$\,mas\, yr$^{-1}$, which~corresponds to an apparent speed of $\beta_\mathrm{app}=  9.4 \pm 5.5$. At~large core separations, component C4 could only be detected at two epochs. Therefore, it is unsuitable for further proper motion calculations. The~core separations for each component and~the fitted linear proper motion curve for C3 are shown in Figure~\ref{fig:SXtav}.

\begin{figure}[H]
\centering
\begin{minipage}[t]{0.47\textwidth}
\centering
\includegraphics[trim={0 2cm 0 1,1cm},clip, width=\textwidth]{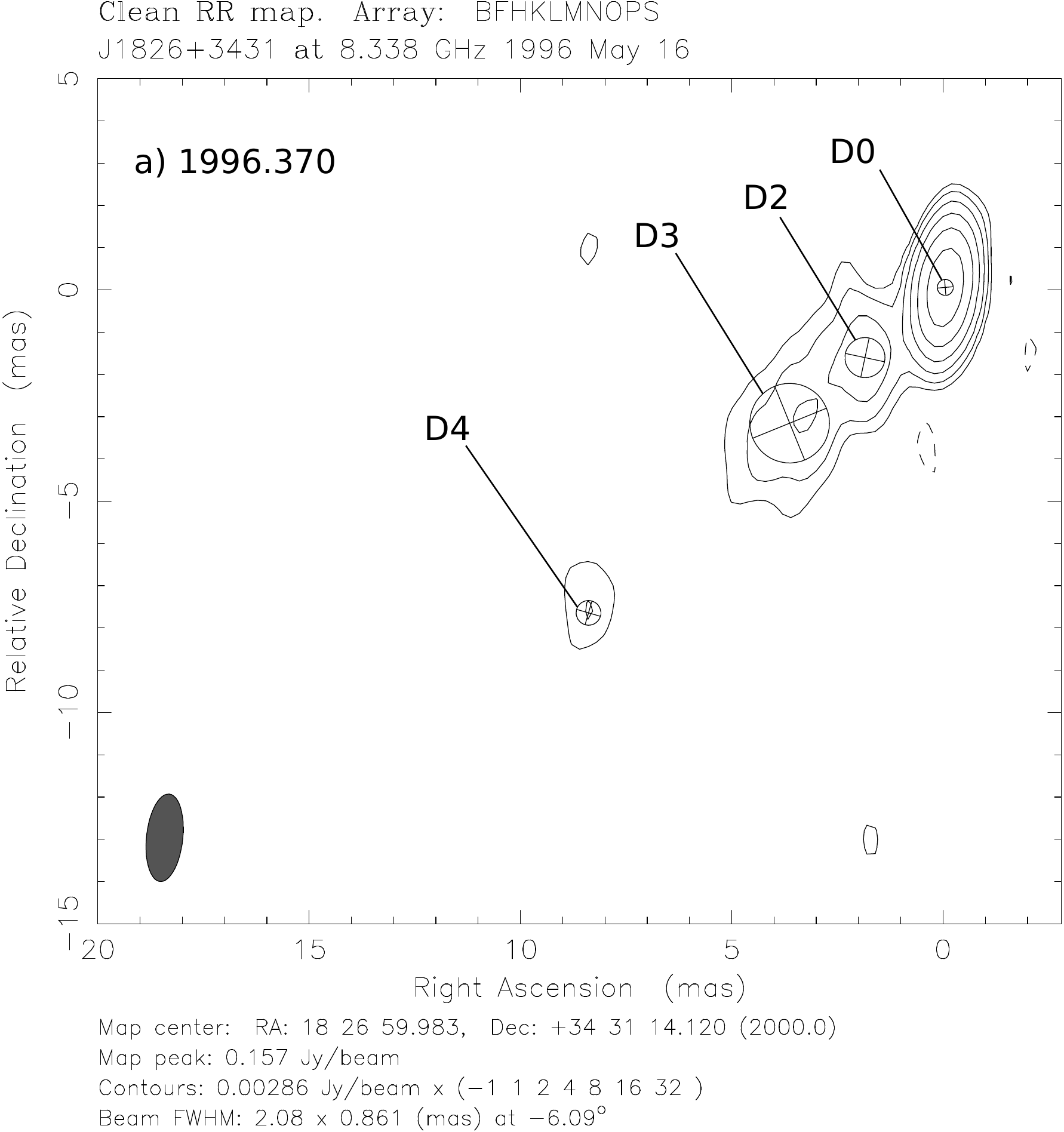}
\end{minipage}
\begin{minipage}[t]{0.47\textwidth}
\centering
\includegraphics[trim={0 2cm 0 1,1cm},clip, width=\textwidth]{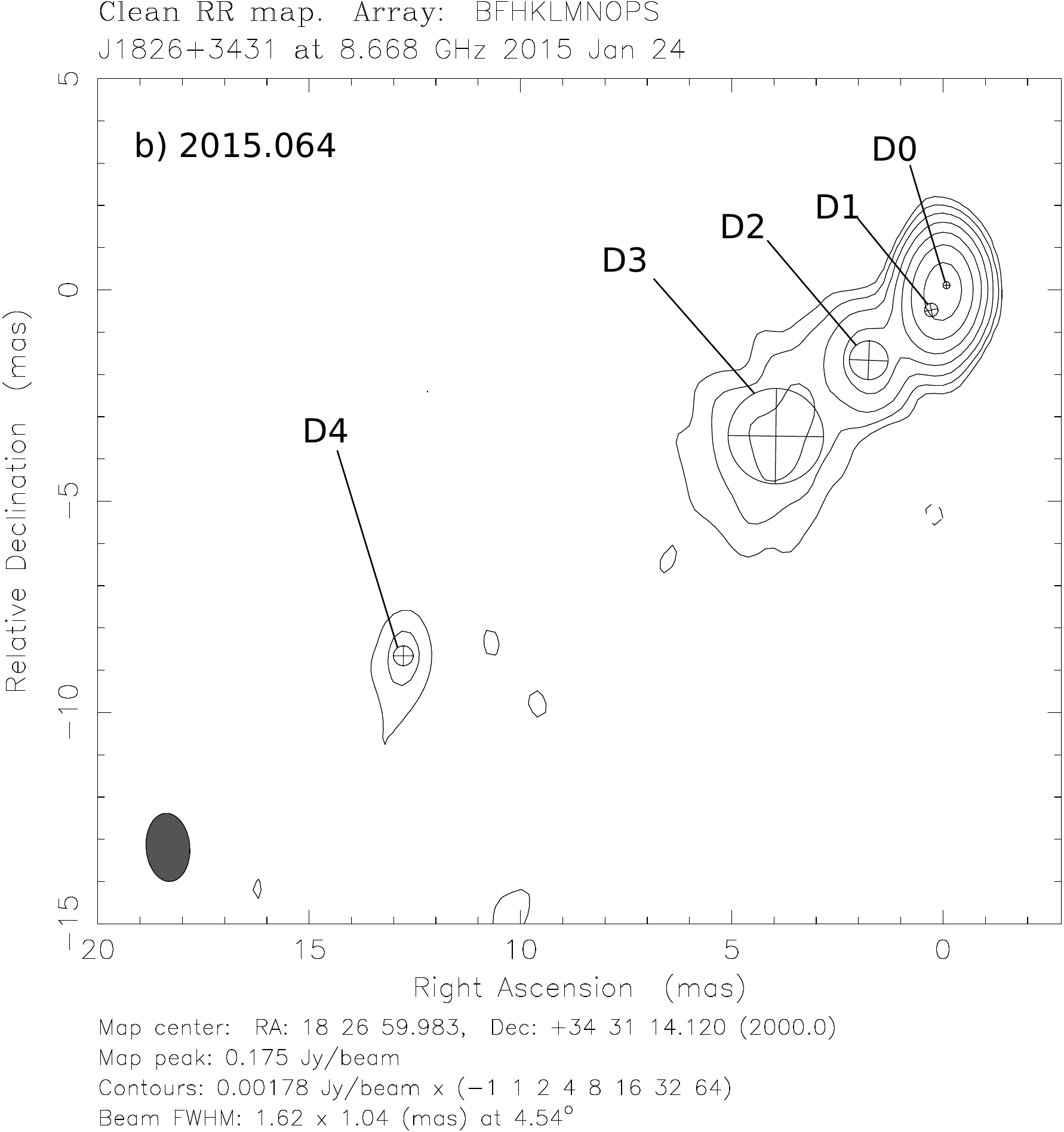}
\end{minipage}
\begin{minipage}[t]{0.47\textwidth}
\centering
\includegraphics[trim={0 2cm 0 1,1cm},clip, width=\textwidth]{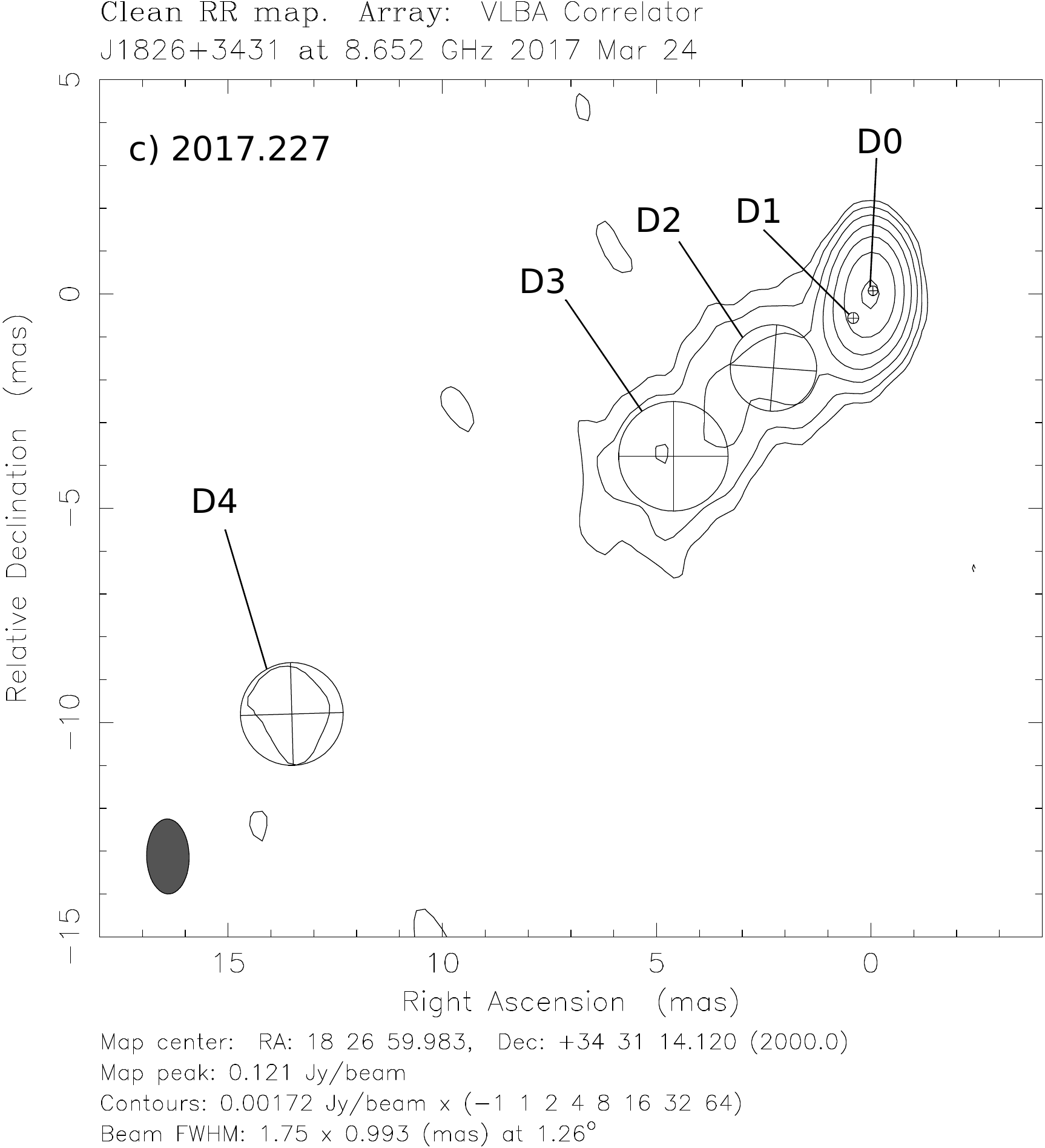}
\end{minipage}
\begin{minipage}[t]{0.47\textwidth}
\centering
\includegraphics[trim={0 2cm 0 1,1cm},clip, width=\textwidth]{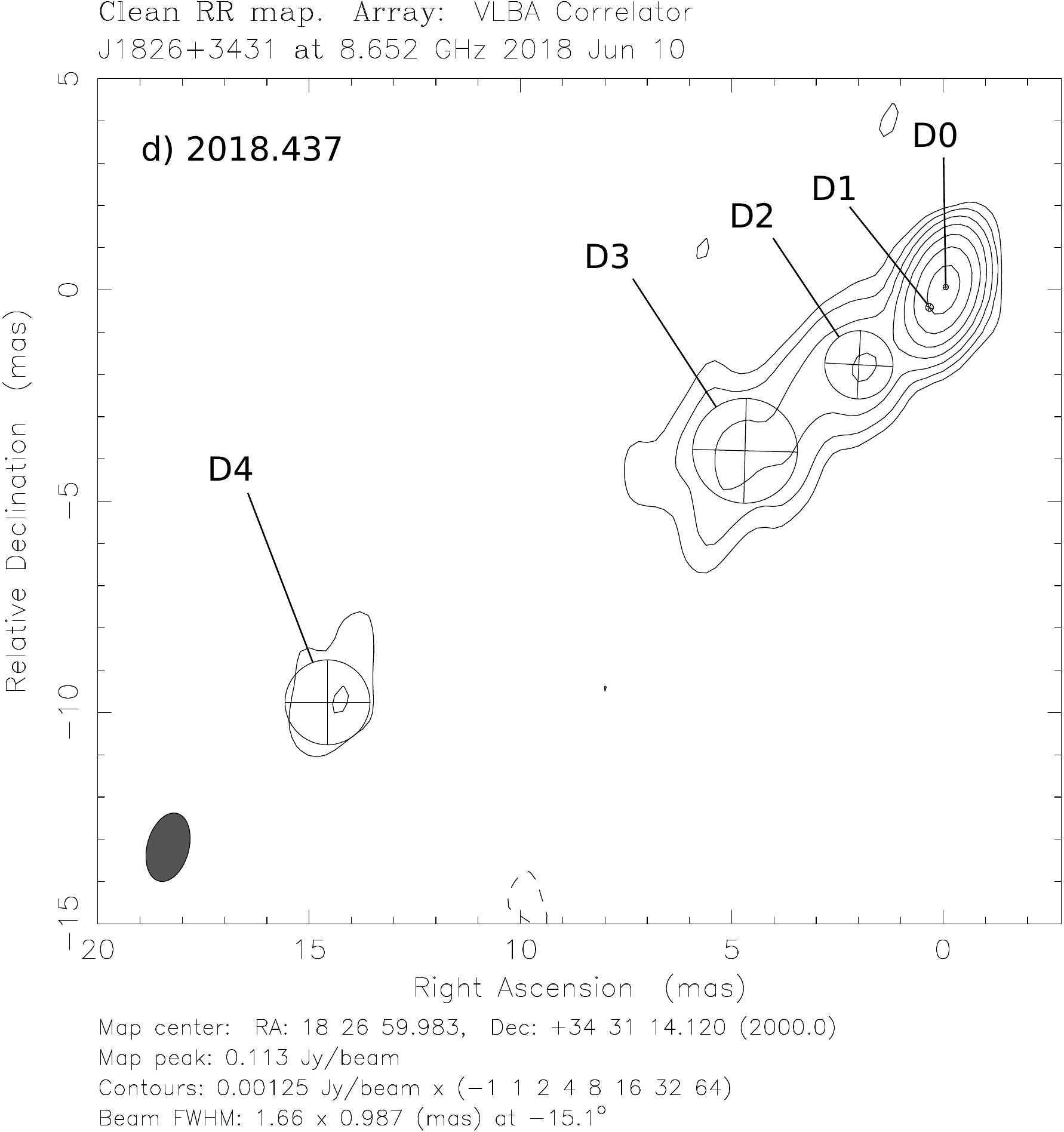}
\end{minipage}
\caption{\label{fig:maps2}Naturally-weighted VLBI images of J1826$+$3431 at epochs (\textbf{a}) 1996.370; (\textbf{b}) 2015.064; (\textbf{c}) 2017.227; and~(\textbf{d}) 2018.437 taken at $8.3/8.7$ GHz restored with the Gaussian model components fitted to the visibility data. The~location and~size (FWHM) of the model components are indicated with circles. In each image, the~elliptical Gaussian restoring beam (FWHM) is shown at the lower left corner. Further details of the images are given in Table~\ref{tab:maps_details2}.}\
\end{figure}

At $8.3/8.7$\,GHz, three jet components, D2, D3, and~D4, can be detected at all four epochs, \mbox{while D1 can only be} detected at the last three epochs. This may indicate that it was ejected some time between the epochs $1996.370$ and~$2015.064$. No discernible proper motions could be detected for components D1 and~D2, while D3 and~D4 show apparent superluminal motions (Table~\ref{app.speeds}). The~core separations and~the fitted linear curves for D3 and~D4 are shown in Figure~\ref{fig:SXtav}. As we performed proper motion and~jet parameter calculations using D3 and~D4, we present contour maps of J1826+3431 made at this frequency at all four epochs in Figure~\ref{fig:maps2}, indicating the fitted model components. 

\unskip
\unskip
\begin{table}[H]
\caption{Parameters describing the movements of the components identified at $8.3/8.7$\,GHz. The~component identification, the~angular proper motion, and~the apparent jet speed in units of $c$ are given in columns 1, 2, and~3, respectively.}
\label{app.speeds}
\centering
\begin{tabular}{ccc}
\toprule
\textbf{ID} & \boldmath{$\mu$}\textbf{ (mas yr$^{-1}$)}  & \boldmath{$\beta_\mathrm{app}$} \\
\midrule 
D3 & $0.06 \pm 0.02$ & $4.8 \pm 1.6$ \\
D4 & $0.28 \pm 0.02$ & $22.2 \pm 1.6$ \\
\bottomrule
\end{tabular}
\end{table}
\unskip
\begin{table}[H]
\caption{\label{tab:maps_details2}Details of the contour maps shown in Figure~\ref{fig:maps2}. The~restoring beam major axis position angle (PA) is measured from north through east. Further positive contour levels in Figure~\ref{fig:maps2} increase by a factor of 2.}
\centering
\begin{tabular}{cccccccc}
\toprule
\vspace{1pt}

 & \textbf{Peak} & \textbf{Image} & \textbf{Lowest} & \multicolumn{3}{c}{\textbf{Restoring Beam}} \\ \cline{5-7}
 \vspace{1pt}
\textbf{Epoch} & \textbf{Brightness} & \textbf{Noise Level} & \textbf{Contour} & \textbf{Major Axis} & \textbf{Minor Axis} & \textbf{PA}\\ 

 & \textbf{(mJy\,beam}\boldmath{$^{-1}$)} & \textbf{(mJy\,beam}\boldmath{$^{-1}$)} & \textbf{(mJy\,beam}\boldmath{$^{-1}$)} & \textbf{(mas)} & \textbf{(mas)} & \textbf{(\boldmath{$^\circ$})} \\
\midrule
$1996.370$& $0.157$ & $0.5$ & $\pm2.8$  & $2.08$ & $0.86$  & $-6.1$ \\
$2015.064$ & $0.175$ & $0.3$ & $\pm1.8$ & $1.62$ & $1.04$  & $4.5$  \\
$2017.227$ & $0.121$ & $0.3$ & $\pm1.7$ & $1.75$ & $0.99$ & $-1.3$  \\
$2018.437$ & $0.113$ & $0.2$ & $\pm1.3$ & $1.66$ & $0.99$  & $-15.1$ \\
\bottomrule
\end{tabular}
\end{table}

At $15.3$\,GHz, only three years elapsed between the first and~last observations. This time span is too short for detecting proper motions in any of the fitted components. Component J1 could be first detected at epoch $2004.616$. This may indicate that J1 was ejected some time between the first and~second epochs. Later, as it moved farther away from the core, this component became detectable even at a lower frequency, $8.7$\,GHz, at the epoch $2015.064$ as D1.

To summarize, apparent superluminal motions could be detected in three components at relatively larger core separations (C3, D3 and~D4) in the jet of J1826$+$3431. Interestingly, the~highest apparent speed is shown by a component (D4) located at the largest core separation ($\sim$10\,mas) implying acceleration in the jet. Acceleration is often detected in the jets of blazars (e.g.,~\cite{Lister_2016}), and~although it usually takes place at smaller core separations, there are examples for faster components further down the jets (e.g.,~PMN\,J0405$-$1308~\cite{Lister_2016,Hervet2016}, PKS 2201$+$171~\cite{Lister_2019}).

\subsection{Relativistic Beaming in J1826$+$3431}

At the highest observing frequency, $15.3$\,GHz, which provides the best angular resolution, we~derived the brightness temperature of the core component using the following equation~\cite{feny.hom}:
\begin{equation}
T _\textrm{b} = 1.22 \cdot 10^{12} (1+z) \frac{S}{\theta^{2} \nu^{2}} \,\, \textrm{K},
\end{equation}
where $S$ is the flux density of the fitted component in Jy, $\theta$ is the FWHM diameter of the circular Gaussian in mas, and~$\nu$ is the observing frequency in GHz. 

In $2003.655$, we were unable to resolve the core region into two components as at the later epochs. As a test, we tried to fit the innermost region with an elliptical Gaussian component and~the resulting fit confirmed that the core was elongated in the direction of component J1 detected at the later epochs. Therefore, J0 and~J1 were blended in $2003.655$, thus the FWHM size and~the flux density are overestimates of the core parameters at this epoch. Since the brightness temperature is inversely proportional to the square of the FWHM size, its derived value can be regarded as a lower limit, $\geq2.8 \times 10^{11}$\,K. In $2004.616$, the~obtained brightness temperature is $T_{\textrm{b}}=(12.3 \pm 1.4) \times 10^{11}$\,K. At~epoch $2005.427$, we could only derive an upper limit for the size of the core component, thus only a lower limit for the brightness temperature can be given, $\geq$12 $\times$ 10$^{11}$\,K. The~lower limits agree with the $T_\mathrm{b}$ estimate made at the middle epoch within its errors. 

\newpage
All brightness temperature values exceed the equipartition limit, $T_\textrm{int} \approx 5\cdot 10^{10}$\,K~\cite{equipartition}, indicating~relativistic beaming in the source. 
The relativistic beaming can be quantified by the Doppler boosting factor, $\delta_{\textrm{vlbi}}$, which can be estimated from the measured $T_\textrm{b}$ value and~the equipartition brightness temperature limit using the following relation~\cite{unified_model}:
\begin{equation}
\delta_{\textrm{vlbi}} = \frac{T_{\textrm{b}}}{T_{\textrm{int}}}.
\end{equation}

The obtained brightness temperature value corresponds to a Doppler factor of  $\delta_{\textrm{vlbi}} = 24.6 \pm 2.9$. 

\subsection{Jet Parameters}

The jet parameters, the~bulk Lorentz factor ($\gamma$), and~the inclination angle with respect to the line of sight ($\phi$) can be calculated from the Doppler factor and~the apparent jet speed as~\cite{unified_model}
\begin{equation}\label{lor3}
\gamma = \frac{\beta_{\textrm{app}}^{2}+\delta^{2}+1}{2\delta}
\end{equation}
and 
\begin{equation}
\tan \phi = \frac{2\beta_{\textrm{app}}}{\beta_{\textrm{app}}^{2}+\delta^{2}-1}.
\end{equation}

The component close to the innermost core region with the best-defined positions and~thus proper motion is D3; therefore, we used the apparent superluminal speed obtained for this jet feature, $\beta_\mathrm{app}^\mathrm{D3}=4.8\pm 1.6$, to derive the jet parameters. Using the Doppler factor obtained from the $15.3$-GHz measurement at epoch $2004.616$, we get a Lorentz factor of $\gamma = 12.8 \pm 1.4$, implying relativistic bulk jet velocity in units of the speed of light, $\beta = 0.997 \pm 0.002$. 
The corresponding jet inclination angle is $\phi = 0.9^{\circ} \pm 0.3^{\circ}$.

\subsection{Radio Spectrum of J1826$+$3431}

\begin{figure}[H]
\centering
 \includegraphics[width=0.6\textwidth, bb=10 40 710 550 , clip=,]{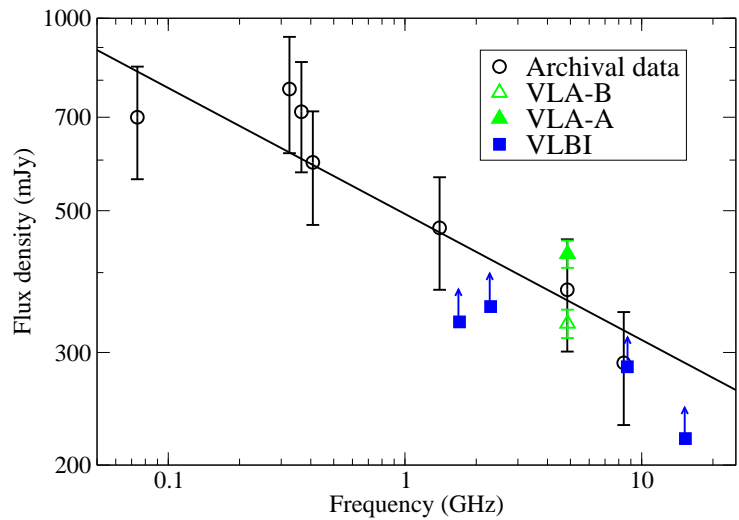}
 \caption{\label{fig:spx2}Radio spectrum of J1826$+$3431. Black circles are from archival measurements, their details are given in Table~\ref{tab:spx}. Green triangles indicate the sum of modelfit components of the VLA measurements presented in this paper. Black line represents the power-law fit to the above total flux density points. Blue squares with arrows show the sum of modelfit components of high-resolution VLBI measurements from this paper, as lower limits to the total flux density. The~selected VLBA epochs ($2015.064$ for $2.3$ and~$8.7$\,GHz, and~$2005.427$ for $15.3$\,GHz) are the closest in time to the $1.7$-GHz EVN measurement.}
\end{figure}

We collected the radio flux density measurements of J1826$+$3431 from the literature (Table~\ref{tab:spx}). We~fitted a power--law curve to these data points to obtain the spectral index ($\alpha$; defined as $S\propto\nu^\alpha$, where $S$ is the flux density and~$\nu$ is the observing frequency). Since the flux density measurements are not simultaneous, source variability may affect the spectral index calculation, so the value can be considered as tentative. The~resulting fit is shown with a black line in Figure~\ref{fig:spx2}, the~obtained spectral index is $\alpha=-0.19 \pm 0.05$, indicating a flat spectrum.

For comparison, we also show the sum of the VLBI Gaussian model component flux densities with blue squares and~upward arrows in Figure~\ref{fig:spx2}. For each observing frequency, we have chosen the observation that is the closest to the $1.7$-GHz EVN measurement. VLBI observations are insensitive to the large (arcsec) scale radio structure; therefore, in the presence of such emission, they underestimate the total flux density of a source. This can be seen in Figure~\ref{fig:spx2} where all VLBI points are indeed below the fitted spectrum.

\begin{table}[H]
\caption{Archival radio flux density measurements of J1826$+$3431.}
\label{tab:spx}
\centering
\tablesize{\small}
\begin{tabular}{cccccc}
\toprule
\textbf{\multirow{2}{*}{Observer Instrument}} & \textbf{Frequency} & \textbf{Flux Density} & \textbf{Flux Density Error} & \textbf{\multirow{2}{*}{Reference}} \\
 & \textbf{(GHz)} & \textbf{(Jy)} & \textbf{(Jy)} & \\
\midrule 
VLA-B, VLA-BnA &  0.074 & 0.700 & 0.140 & VLSS~\cite{VLSS} \\
Westerbork Synthesis Radio Telescope & 0.325 & 0.775 & 0.160 & WN~\cite{WN} \\
Texas Interferometer & 0.365 & 0.714 & 0.140 & TXS~\cite{TXS}\\
Bologna Northern Cross Telescope & 0.408 & 0.595 & 0.120 & B2.3~\cite{B2.3} \\
VLA-D, VLA-DnC & 1.4 & 0.470 & 0.094 & NVSS~\cite{nvss} \\
 Green Bank 91-m telescope & 4.85 & 0.376 & 0.075 & 87GB~\cite{87GB} \\
 VLA-A & 8.4 & 0.289 & 0.058 & CLASS~\cite{CLASS} \\
\bottomrule
\end{tabular}
\end{table}
\unskip

\begin{figure}[H]
\centering
 \includegraphics[width=0.6\textwidth]{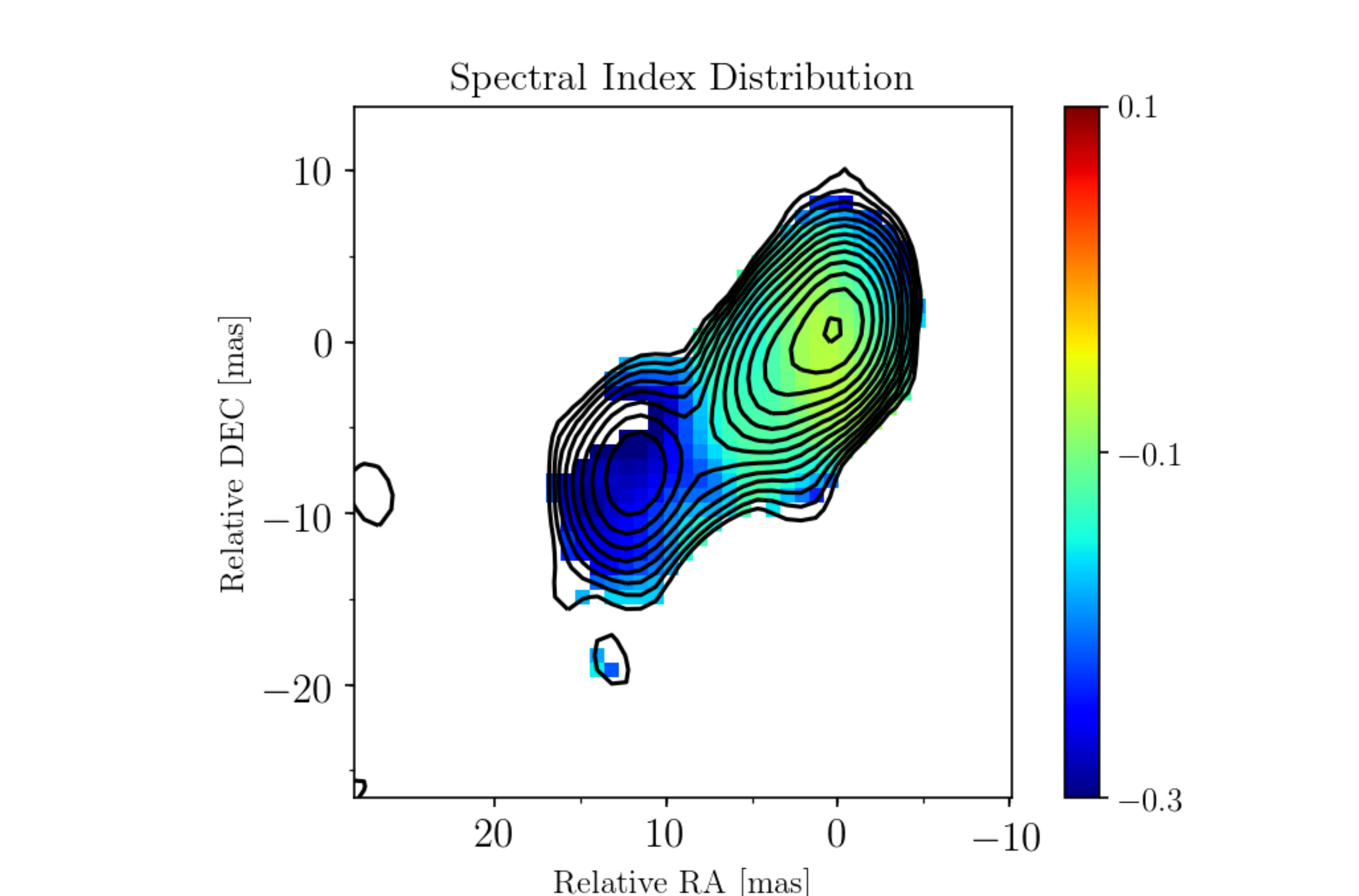}
 \caption{\label{fig:spx1}Spectral index map of J1826$+$3431 using $2.3$- and~$8.7$-GHz data taken at 2015.064. The~colour bar shows the spectral index values. The~contours represent the 2.3-GHz image, the~lowest contour level is drawn at $3\sigma$ image noise level corresponding to 2.4~mJy\,beam$^{-1}$. Further contours increase by a factor of 2. The~peak intensity is 162~mJy\,beam$^{-1}$. The~restoring beam size is $3.69\mathrm{\,mas} \times 5.72 \mathrm{\,mas}$ at $\mathrm{PA}=-3.7^\circ$.}
\end{figure}

The simultaneous $2.3$- and~$8.3/8.7$-GHz observations allow us to map the spectral index distribution of the mas-scale radio emission in J1826$+$3431 using VIMAP~\cite{vimap}. We~present a spectral index image in Figure~\ref{fig:spx1}. We~used our data from the epoch 2015.064. After using elliptical masks to exclude the cores from the images, we matched the $2.3$- and~$8.7$-GHz maps of J1826$+$3431 by using two-dimensional correlation. Considering that the frequencies are relatively close to each other and~that the measurements were performed at the same time, it was not necessary to shift the brightness peak. The~spectral index image (Figure~\ref{fig:spx1}) suggests a flat-spectrum radio emission ($\alpha \approx  -0.1 \pm 0.1$) in the core region and~along the jet closer to the core. It is in good agreement with the spectral index calculated using the archival total flux density data (Figure~\ref{fig:spx2}) since the dominant emission feature of the source is the core. The~spectrum gradually steepens further away from the core along the jet, as expected from optically thin synchrotron radio emission.

\subsection{Flux Density Variability}

\begin{figure}[H]
\centering
 \includegraphics[width=0.7\textwidth]{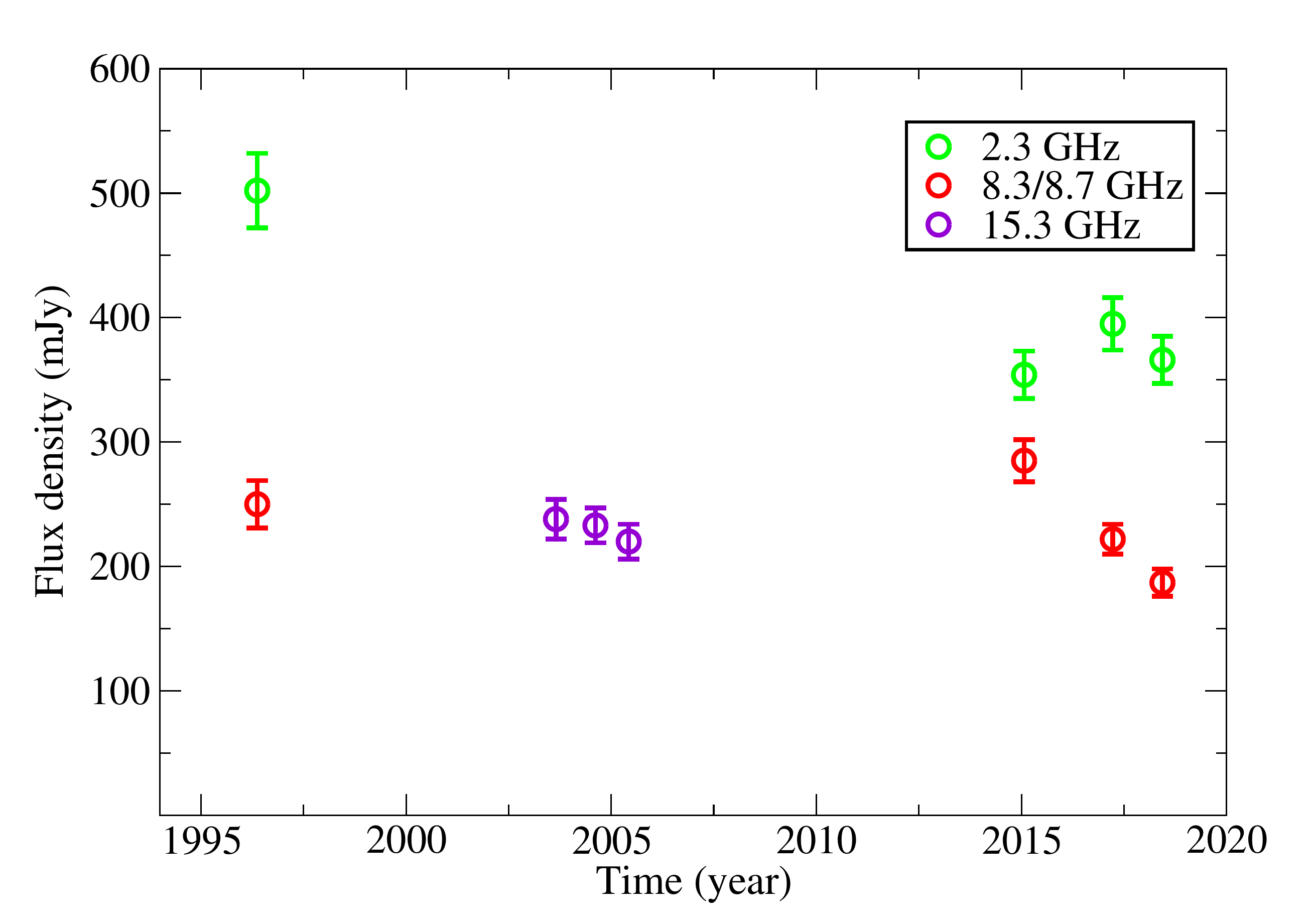}
 \caption{\label{fluxdensity}Flux density of J1826+3431 versus time at $2.3$, $8.3/8.7$, and~15.3\,GHz, based on the sum of fitted VLBI model component flux densities.}
\end{figure}

To characterize variability in the VLBI-detected features, we plot the sum of flux densities in the fitted Gaussian model components at different observing frequencies as a function of time in Figure~\ref{fluxdensity}. As can be clearly seen, especially for the 2.3- and~8.3/8.7-GHz data where the observations span more than two decades, the~flux density of the source is strongly variable over time. Moreover, the~variations in the difference of the 2.3- and~8.3/8.7-GHz flux density values suggest variability \mbox{of the spectral index.}

\subsection{$\gamma$-Ray Properties of 3EG\,J1824$+$3441}

3EG\,J1824$+$3441 was detected with a weak $\gamma$-ray flux of $(28.7 \pm 9.3) \times 10^{-8}$\,photon\,s$^{-2}$\,cm$^{-1}$ by EGRET~\cite{3EG}. However, an alternate version of the EGRET catalogue~\cite{alternate_3EG} (see also~\cite{Thompson2008}) which used reprocessed data assuming a different model for the Galactic interstellar emission does not contain as many as 107 sources previously detected in~\cite{3EG}, including also 3EG\,J1824$+$3441. In that sense, it~is not surprising that no $\gamma$-ray source is listed in the most recent fourth {\it Fermi}/LAT catalogue~\cite{4FGL_cat} \mbox{within a $1^\circ$~radius} of the position of the putative EGRET $\gamma$-ray source.

Alternatively, even if the source exists, long-term variability of its $\gamma$-ray flux can also explain the {\it Fermi} non-detection. Indeed, a comparison of high-confidence ($>$4$\sigma$) EGRET detections with the third {\it Fermi}/LAT catalogue revealed $10$ extragalactic $\gamma$-ray sources missing from the latter~\cite{egret_fermi_var_2017}. Subsequent analysis of seven years of {\it Fermi} data showed that five sources were in a low-luminosity state during the {\it Fermi} observations compared to their luminosity during the EGRET observations~\cite{egret_fermi_var_2017}. The~generally decreasing long-term trend we see in the radio flux density of the possible counterpart, the~blazar J1826+3431, with respect to the 1990s when CGRO was operational (see e.g., our VLA results and~Figure~\ref{fluxdensity}) is also consistent with a $\gamma$-ray flux decrease.

\section{Conclusions}
We analyzed eight epochs of high-resolution multi-frequency VLBI data of the radio source J1826$+$3431, which is the proposed radio counterpart of the EGRET $\gamma$-ray source, 3EG\,J1824$+$3441. The~archival observations span a period of more than 22 years. 
Each data set was imaged, and~the brightness distribution of the source was modeled using circular Gaussian components, in order to characterize the physical and~geometric properties of the jet and~to analyze the kinematics of J1826$+$3431. One and~two moving jet components could be securely identified at multiple epochs in the $2.3$- and~$8.3/8.7$-GHz data, respectively. All of them showed outward motion with apparent superluminal speed. The~estimated brightness temperature of the most compact component, the~core at $15.3$\,GHz suggests that the radio emission is relativistically beamed in the source. 
From the relativistic effects seen in the radio jet and~the strong variability of the core component at the measured frequencies, the~radio source J1826$+$3431 can be identified as a blazar. Since the vast majority of extragalactic $\gamma$-ray sources are associated with blazars, we can conclude that J1826$+$3431 is very likely the radio counterpart of the unidentified EGRET source 3EG\,J1824$+$3441. However, no $\gamma$-ray source is found at this position in the latest {\it Fermi}/LAT catalogue~\cite{4FGL_cat}, and~its detection can be regarded uncertain based on an alternative version of the EGRET catalogue~\cite{alternate_3EG}.

\vspace{6pt} 

\authorcontributions{Conceptualization and~investigation, K.\'E.G. and~P.M.V. Formal analysis P.M.V. Writing---original draft preparation, P.M.V. and~K.\'E.G. Writing---review and~editing, K.\'E.G. and~S.F. All authors have read and~agreed to the published version of the manuscript.}

\funding{
For this research, K.\'E.G. received funding from the J\'anos Bolyai Research Scholarship of the Hungarian Academy of Sciences and~from the \'UNKP-19-4 New National Excellence Program of the Ministry of For Innovation and~Technology. {This work was supported by the Hungarian Research, Development and~Innovation Office (OTKA K134213).}}

\acknowledgments{The National Radio Astronomy Observatory is a facility of the National Science Foundation operated under cooperative agreement by Associated Universities, Inc. The~European VLBI Network is a joint facility of independent European, African, Asian, and~North American radio astronomy institutes. Scientific results from data presented in this publication are derived from the following EVN project code: EF025. We~acknowledge the use of archival calibrated VLBI data from the Astrogeo Center database maintained by Leonid Petrov.}

\conflictsofinterest{The authors declare no conflict of interest. The~funders had no role in the design of the study; in the collection, analyses, or interpretation of data; in the writing of the manuscript, or in the decision to publish the results.}

\abbreviations{The following abbreviations are used in this manuscript:\\

\noindent 
\begin{tabular}{@{}ll}
AGN & Active Galactic Nuclei \\
CGRO & Compton Gamma Ray Observatory \\
EGRET & Energetic Gamma Ray Experiment Telescope \\
EVN & European VLBI Network \\
FWHM & Full Width at Half-Maximum \\
NRAO & U.S. National Radio Astronomy Observatory \\
$\Lambda$CDM & Lambda Cold Dark Matter \\
LAT & Large Area Telescope \\
NVSS & NRAO VLA Sky Survey \\
PA & Position Angle \\
VLA & Karl G. Jansky Very Large Array \\
VLBA & Very Long Baseline Array \\
VLBI & Very Long Baseline Interferometry \\
SED & Spectral Energy Distribution \\
SNR & Signal-to-Noise Ratio 
\end{tabular}}

\appendixtitles{no} 
\appendix
\section{}
Here, we provide the results of the model fitting to the VLBI visibility data. In Tables \ref{tabLk}--\ref{tabUk}, we list the best-fit parameters of the circular Gaussian components used to describe the brightness distribution of the $1.7$-, $2.3$-, $8.3/8.7$-, and~$15.3$-GHz calibrated data of J1826$+$3431, respectively.

\begin{table}[H]
\caption{Parameters of the fitted Gaussian model components at $1.7$\,GHz. The~observational epoch is given in column 1. The~flux densities, relative positions in right ascension and~declination directions with respect to the core component, and~the FWHM sizes are listed in columns 2, 3, 4, and~5, respectively. In the final column we give the component identifiers.}
\label{tabLk}
\centering
\tablesize{\normalsize}
\begin{tabular}{cccccc}
\toprule
\textbf{Epoch} & \textbf{Flux Density} & \textbf{$\Delta \alpha$} & \textbf{$\Delta \delta$} & \textbf{FWHM} &\textbf{\multirow{2}{*}{ID}} \\
\textbf{(Year)} & \textbf{(Jy)} & \textbf{(mas)} & \textbf{(mas)} & \textbf{(mas)} & \\
\midrule 
\multirow{3}{*}{2014.145} & 0.236 $\pm$ 0.024 & 0 & 0 & $3.83 \pm 0.01$ &  L0 \\ 
& 0.081 $\pm$ 0.008 & 9.78 $\pm$ 1.91 & $-7.13 \pm 1.91$ & 4.07 $\pm$ 0.04 & L1 \\
& 0.018 $\pm$ 0.002 & 22.82 $\pm$ 1.92 & $-22.90 \pm 1.92$ & 7.66 $\pm$ 0.32 & L2 \\
\bottomrule
\end{tabular}
\end{table}
\unskip
\begin{table}[H]
\caption{Parameters of the fitted Gaussian model components at $2.3$\,GHz. The~observational epochs are given in column 1. The~flux densities, relative positions in right ascension and~declination directions with respect to the core component, and~the FWHM sizes are listed in columns 2, 3, 4, and~5, respectively. In the final column, we give the component identifiers. Asterisks (*) denote those epochs where we time-averaged the calibrated data into $10$-s blocks.} 
\label{tabSk}
\centering
\tablesize{\normalsize}
\begin{tabular}{cccccc}
\toprule
\textbf{Epoch} & \textbf{Flux Density} & \boldmath{$\Delta \alpha$} & \boldmath{$\Delta \delta$} & \textbf{FWHM} &\textbf{\multirow{2}{*}{ID}} \\
\textbf{(Year)} & \textbf{(Jy)} & \textbf{(mas)} & \textbf{(mas)} & \textbf{(mas)} & \\
\midrule 
\multirow{4}{*}{1996.370 *} & 0.228 $\pm$ 0.023 & 0 & 0 & $0.96 \pm 0.01$ & C0 \\
& 0.155 $\pm$ 0.016 & 2.87 $\pm$ 1.11 & $-2.60 \pm 1.11$ & 2.07 $\pm$ 0.02 & Ca \\
& 0.103 $\pm$ 0.010 & 10.62 $\pm$ 1.12 & $-7.67 \pm 1.12$ & 5.34 $\pm$ 0.33 & C3 \\
& 0.016 $\pm$ 0.002 & 23.95 $\pm$ 1.21 & $-24.36 \pm 1.21$ & 5.98 $\pm$ 0.79 & C4 \\
\midrule
\multirow{4}{*}{2015.064 *} & 0.120 $\pm$ 0.012 & 0 & 0 & $0.93 \pm 0.01$ & C0 \\
& 0.110 $\pm$ 0.011 & 2.26 $\pm$ 1.03 & $-1.98 \pm 1.03$ & 2.36 $\pm$ 0.01 & C1 \\
& 0.059 $\pm$ 0.006 & 4.90 $\pm$ 1.03 & $-4.23 \pm 1.03$ & 2.00 $\pm$ 0.02 & C2 \\
& 0.065 $\pm$ 0.007 & 12.31 $\pm$ 1.03 & $-9.08 \pm 1.03$ & 3.92 $\pm$ 0.08 & C3 \\
\midrule
\multirow{4}{*}{2017.227} & 0.146 $\pm$ 0.015 & 0 & 0 & $0.69 \pm 0.01$ & C0 \\
& 0.099 $\pm$ 0.010 & 2.65 $\pm$ 1.09 & $-2.06 \pm  1.09$ & 2.06 $\pm$ 0.01 & C1 \\
& 0.070 $\pm$ 0.007 & 5.15 $\pm$ 1.09 & $-4.29 \pm 1.09$ & 2.29 $\pm$ 0.02 & C2 \\
& 0.080 $\pm$ 0.008 & 12.59 $\pm$ 1.09 & $-9.16 \pm 1.09$ & 4.21 $\pm$ 0.12 & C3 \\
\midrule
\multirow{5}{*}{2018.437} & 0.138 $\pm$ 0.014 & 0 & 0 & $0.22 \pm 0.01$ & C0 \\
& 0.073 $\pm$ 0.007 & 2.65 $\pm$ 1.05 & $-1.93 \pm 1.05$ & 1.19 $\pm$ 0.03 & C1 \\
& 0.071 $\pm$ 0.007 & 5.05 $\pm$ 1.05 & $-4.15 \pm 1.05$ & 2.34 $\pm$ 0.08 & C2 \\
& 0.072 $\pm$ 0.007 & 12.67 $\pm$ 1.06 & $-9.29 \pm 1.06$ & 3.89 $\pm$ 0.22 & C3 \\
& 0.012 $\pm$ 0.001 & 26.50 $\pm$ 1.13 & $-25.56 \pm 1.13$ & 4.81 $\pm$ 1.06 & C4 \\
\bottomrule 
\end{tabular}
\end{table}
\unskip
\begin{table}[H]
\caption{Parameters of the fitted Gaussian model components at $8.3/8.7$\,GHz. The~observational epochs are given in column 1. The~flux densities, relative positions in right ascension and~declination directions with respect to the core component, and~the FWHM sizes are listed in columns 2, 3, 4, and~5, respectively. In the final column, we give the component identifiers. Asterisks (*) denote those epochs where we time-averaged the calibrated data into $10$-s blocks.} 
\label{tabXk}
\centering
\tablesize{\normalsize}
\begin{tabular}{cccccc}
\toprule
\textbf{Epoch} & \textbf{Flux Density} & \boldmath{$\Delta \alpha$} & \boldmath{$\Delta \delta$} & \textbf{FWHM} &\textbf{\multirow{2}{*}{ID}} \\
\textbf{(Year)} &\textbf{ (Jy)} & \textbf{(mas)} & \textbf{(mas)} & \textbf{(mas)} & \\
\midrule 
\multirow{4}{*}{1996.370 *} & 0.175 $\pm$ 0.018 & 0 & 0 & $0.38 \pm 0.01$ & D0 \\
& 0.031 $\pm$ 0.003 & 1.90 $\pm$ 0.30 & $-1.66 \pm 0.30$ & 0.95 $\pm$ 0.07 & D2 \\
& 0.036 $\pm$ 0.004 & 3.68 $\pm$ 0.30 & $-3.21 \pm 0.30$ & 1.88 $\pm$ 0.03 & D3 \\
& 0.008 $\pm$ 0.001 & 8.44 $\pm$ 0.30 & $-7.70 \pm 0.30$ & 0.58 $\pm$ 0.08 & D4 \\
\midrule 
\multirow{4}{*}{2015.064 *} & 0.142 $\pm$ 0.014 & 0 & 0 & $0.17 \pm 0.01$ & D0 \\
& 0.067 $\pm$ 0.007 & 0.36 $\pm$ 0.29 & $-0.59 \pm 0.29$ & 0.32 $\pm$ 0.01 & D1  \\
& 0.035 $\pm$ 0.004 & 1.84 $\pm$ 0.29 & $-1.77 \pm 0.29$  & 0.92 $\pm$ 0.02 & D2 \\
& 0.036 $\pm$ 0.004 & 4.04 $\pm$ 0.42 & $-3.57 \pm 0.42$ & 2.26 $\pm$ 0.22 & D3 \\
& 0.005 $\pm$ 0.001 & 12.85 $\pm$ 0.29 & $-8.77 \pm 0.29$ & 0.47 $\pm$ 0.03 & D4 \\
\midrule 
\multirow{4}{*}{2017.227} & 0.111 $\pm$ 0.011 & 0 & 0 & $0.22 \pm 0.01$ & D0 \\
& 0.032 $\pm$ 0.003 & 0.46 $\pm$ 0.30 & $-0.64 \pm 0.30$ & 0.26 $\pm$ 0.01 & D1  \\
& 0.032 $\pm$ 0.003 & 2.32 $\pm$ 0.31 & $-1.80 \pm 0.31$ & 2.02 $\pm$ 0.09 & D2 \\
& 0.037 $\pm$ 0.004 & 4.66 $\pm$ 0.33 & $-3.86 \pm 0.33$ & 2.56 $\pm$ 0.31 & D3  \\
& 0.010 $\pm$ 0.001 & 13.56 $\pm$ 0.34 & $-9.88 \pm 0.34$ & 2.40 $\pm$ 0.33 & D4 \\
\midrule 
\multirow{5}{*}{2018.437} & 0.099 $\pm$ 0.010 & 0 & 0 & $0.13 \pm 0.01$ & D0 \\
& 0.021 $\pm$ 0.002 & 0.38 $\pm$ 0.29 & $-0.48 \pm 0.29$ & 0.18 $\pm$ 0.01 & D1  \\
& 0.028 $\pm$ 0.003 & 2.05 $\pm$ 0.29 & $-1.84 \pm 0.29$ & 1.62 $\pm$ 0.02 & D2 \\
& 0.031 $\pm$ 0.003 & 4.75 $\pm$ 0.32 & $-3.87 \pm 0.32$ & 2.48 $\pm$ 0.07 & D3  \\
& 0.008 $\pm$ 0.001 & 14.63 $\pm$ 0.33 & $-9.83 \pm 0.33$ & 2.01 $\pm$ 0.10 & D4 \\
\bottomrule
\end{tabular}
\end{table}
\unskip
\begin{table}[H]
\caption{Parameters of the fitted Gaussian model components at $15.3$\,GHz. The~observational epochs are given in column 1. The~flux densities, relative positions in right ascension and~declination directions with respect to the core component, and~the FWHM sizes are listed in columns 2, 3, 4, and~5, respectively. In the final column, we give the component identifiers.} 
\label{tabUk}
\centering
\begin{tabular}{cccccc}
\toprule
\textbf{Epoch} & \textbf{Flux Density} & \boldmath{$\Delta \alpha$} & \boldmath{$\Delta \delta$} & \textbf{FWHM} &\textbf{\multirow{2}{*}{ID}} \\
\textbf{(year)} & \textbf{(Jy)} & \textbf{(mas)} & \textbf{(mas)} & \textbf{(mas)} & \\
\midrule 
\multirow{5}{*}{2003.655} & 0.154 $\pm$ 0.015 & 0 & 0 & $0.085 \pm 0.001$ & J0 \\
& 0.021 $\pm$ 0.002 & 0.58 $\pm$ 0.16 & $-0.75 \pm 0.16$ & 0.331 $\pm$ 0.007 & J2 \\
& 0.031 $\pm$ 0.003 & 1.93 $\pm$ 0.16 & $-1.62 \pm 0.16$ & 0.358 $\pm$ 0.003 & J3 \\
& 0.020 $\pm$ 0.002 & 3.23 $\pm$ 0.17 & $-2.43 \pm 0.17$ & 1.270 $\pm$ 0.148 & J4 \\
& 0.012 $\pm$ 0.001 & 4.54 $\pm$ 0.19 & $-4.13 \pm 0.19$ & 1.939 $\pm$ 0.252 & J5 \\
\midrule 
\multirow{6}{*}{2004.616} & 0.116 $\pm$ 0.012 & 0 & 0 & $0.037 \pm 0.001$ & J0 \\
& 0.036 $\pm$ 0.004 & 0.08 $\pm$ 0.16 & $-0.10 \pm 0.16$ & 0.057 $\pm$ 0.001 & J1 \\
& 0.019 $\pm$ 0.002 & 0.56 $\pm$ 0.16 & $-0.69 \pm 0.16$ & 0.268 $\pm$ 0.004 & J2 \\
& 0.031 $\pm$ 0.003 & 1.93 $\pm$ 0.17 & $-1.59 \pm 0.17$ & 0.439 $\pm$ 0.010 & J3 \\
& 0.025 $\pm$ 0.003 & 3.54 $\pm$ 0.18 & $-2.78 \pm 0.18$ & 1.489 $\pm$ 0.133 & J4 \\
& 0.006 $\pm$ 0.001 & 5.63 $\pm$ 0.28 & $-4.69 \pm 0.28$ & 1.871 $\pm$ 0.441 & J5 \\
\midrule 
\multirow{6}{*}{2005.427} & 0.133 $\pm$ 0.013 & 0 & 0 & $<0.04$ & J0 \\
& 0.017 $\pm$ 0.002 & 0.27 $\pm$ 0.16 & $-0.29 \pm 0.16$ & 0.199 $\pm 0.001$ & J1 \\
& 0.010 $\pm$ 0.001 & 0.78 $\pm$ 0.16 & $-1.04 \pm 0.16$ & 0.320 $\pm$ 0.002 & J2 \\
& 0.033 $\pm$ 0.003 & 1.99 $\pm$ 0.16 & $-1.65 \pm 0.16$ & 0.475 $\pm$ 0.002 &  J3 \\
& 0.023 $\pm$ 0.002 & 3.45 $\pm$ 0.18 & $-2.71 \pm 0.18$ & 1.704 $\pm$ 0.028 & J4 \\
& 0.004 $\pm$ 0.001 & 5.29 $\pm$ 0.17 & $-4.83 \pm 0.17$ & 0.685 $\pm$ 0.020 & J5 \\
\bottomrule
\end{tabular}
\end{table}




\reftitle{References}




\begin{thebibliography}{999}

\bibitem[{Urry} and~{Padovani}(1995)]{unified_model}
{Urry}, C.M.; {Padovani}, P.
\newblock {Unified Schemes for Radio-Loud Active Galactic Nuclei}.
\newblock {\em Publ. Astron. Soc. Pac.} {\bf
  1995}, {\em 107},~803,
\newblock
  doi:{\changeurlcolor{black}\href{https://doi.org/10.1086/133630}{\detokenize{10.1086/133630}}}.

\bibitem[{Giommi} \em{et~al.}(2012){Giommi}, {Padovani}, {Polenta},
  {Turriziani}, {D'Elia}, and~{Piranomonte}]{Giommi2012}
{Giommi}, P.; {Padovani}, P.; {Polenta}, G.; {Turriziani}, S.; {D'Elia}, V.;
  {Piranomonte}, S.
\newblock {A simplified view of blazars: clearing the fog around long-standing
  selection effects}.
\newblock {\em Mon. Not.  R. Astron. Soc.} {\bf 2012},
  {\em 420},~2899--2911,
\newblock
  doi:{\changeurlcolor{black}\href{https://doi.org/10.1111/j.1365-2966.2011.20044.x}{\detokenize{10.1111/j.1365-2966.2011.20044.x}}}.

\bibitem[{Abdollahi} \em{et~al.}(2020){Abdollahi}, {Acero}, {Ackermann},
  {Ajello}, {Atwood}, {Axelsson}, {Baldini}, {Ballet}, {Barbiellini},
  {Bastieri}, {Becerra Gonzalez}, {Bellazzini}, {Berretta}, {Bissaldi},
  {Blandford}, {Bloom}, {Bonino}, {Bottacini}, {Brandt}, {Bregeon}, {Bruel},
  {Buehler}, {Burnett}, {Buson}, {Cameron}, {Caputo}, {Caraveo}, {Casandjian},
  {Castro}, {Cavazzuti}, {Charles}, {Chaty}, {Chen}, {Cheung}, {Chiaro},
  {Ciprini}, {Cohen-Tanugi}, {Cominsky}, {Coronado-Bl{\'a}zquez}, {Costantin},
  {Cuoco}, {Cutini}, {D'Ammando}, {DeKlotz}, {Torre Luque}, {de Palma},
  {Desai}, {Digel}, {Lalla}, {Mauro}, {Venere}, {Dom{\'\i}nguez}, {Dumora},
  {Dirirsa}, {Fegan}, {Ferrara}, {Franckowiak}, {Fukazawa}, {Funk}, {Fusco},
  {Gargano}, {Gasparrini}, {Giglietto}, {Giommi}, {Giordano}, {Giroletti},
  {Glanzman}, {Green}, {Grenier}, {Griffin}, {Grondin}, {Grove}, {Guiriec},
  {Harding}, {Hayashi}, {Hays}, {Hewitt}, {Horan}, {J{\'o}hannesson},
  {Johnson}, {Kamae}, {Kerr}, {Kocevski}, {Kova{\v{c}}evi{\'c}}, {Kuss},
  {Landriu}, {Larsson}, {Latronico}, {Lemoine-Goumard}, {Li}, {Liodakis},
  {Longo}, {Loparco}, {Lott}, {Lovellette}, {Lubrano}, {Madejski}, {Maldera},
  {Malyshev}, {Manfreda}, {Marchesini}, {Marcotulli}, {Mart{\'\i}-Devesa},
  {Martin}, {Massaro}, {Mazziotta}, {McEnery}, {Mereu}, {Meyer}, {Michelson},
  {Mirabal}, {Mizuno}, {Monzani}, {Morselli}, {Moskalenko}, {Negro}, {Nuss},
  {Ojha}, {Omodei}, {Orienti}, {Orlando}, {Ormes}, {Palatiello}, {Paliya},
  {Paneque}, {Pei}, {Pe{\~n}a-Herazo}, {Perkins}, {Persic}, {Pesce-Rollins},
  {Petrosian}, {Petrov}, {Piron}, {Poon}, {Porter}, {Principe}, {Rain{\`o}},
  {Rando}, {Razzano}, {Razzaque}, {Reimer}, {Reimer}, {Remy}, {Reposeur},
  {Romani}, {Parkinson}, {Schinzel}, {Serini}, {Sgr{\`o}}, {Siskind}, {Smith},
  {Spandre}, {Spinelli}, {Strong}, {Suson}, {Tajima}, {Takahashi}, {Tak},
  {Thayer}, {Thompson}, {Tibaldo}, {Torres}, {Torresi}, {Valverde}, {Klaveren},
  {Zyl}, {Wood}, {Yassine}, and~{Zaharijas}]{4FGL_cat}
{Abdollahi}, S.; {Acero}, F.; {Ackermann}, M.; {Ajello}, M.; {Atwood}, W.B.;
  {Axelsson}, M.; {Baldini}, L.; {Ballet}, J.; {Barbiellini}, G.; {Bastieri},
  D.; et al.
\newblock {Fermi Large Area Telescope Fourth Source Catalog}.
\newblock {\em Astrophys. J. Suppl. Ser.} {\bf 2020}, {\em
  247},~33.
\newblock
  doi:{\changeurlcolor{black}\href{https://doi.org/10.3847/1538-4365/ab6bcb}{\detokenize{10.3847/1538-4365/ab6bcb}}}.

\bibitem[{Rani} \em{et~al.}(2016){Rani}, {Krichbaum}, {Hodgson}, and
  {Zensus}]{hadronic}
{Rani}, B.; {Krichbaum}, T.P.; {Hodgson}, J.A.; {Zensus}, J.A.
\newblock {Location and~origin of gamma-rays in blazars}.
\newblock {\em J. Phys. Conf. Ser.} {\bf 2016}, {\em
  718},~052032,
\newblock
  doi:{\changeurlcolor{black}\href{https://doi.org/10.1088/1742-6596/718/5/052032}{\detokenize{10.1088/1742-6596/718/5/052032}}}.

\bibitem[{Dermer}(1995)]{ssc}
{Dermer}, C.D.
\newblock {On the Beaming Statistics of Gamma-Ray Sources}.
\newblock {\em Astrophys. J. Lett.} {\bf 1995}, {\em 446},~L63.
\newblock
  doi:{\changeurlcolor{black}\href{https://doi.org/10.1086/187931}{\detokenize{10.1086/187931}}}.

\bibitem[{Hartman} \em{et~al.}(1999){Hartman}, {Bertsch}, {Bloom}, {Chen},
  {Deines-Jones}, {Esposito}, {Fichtel}, {Friedlander}, {Hunter}, {McDonald},
  {Sreekumar}, {Thompson}, {Jones}, {Lin}, {Michelson}, {Nolan}, {Tompkins},
  {Kanbach}, {Mayer-Hasselwander}, {M{\"u}cke}, {Pohl}, {Reimer}, {Kniffen},
  {Schneid}, {von Montigny}, {Mukherjee}, and~{Dingus}]{3EG}
{Hartman}, R.C.; {Bertsch}, D.L.; {Bloom}, S.D.; {Chen}, A.W.; {Deines-Jones},
  P.; {Esposito}, J.A.; {Fichtel}, C.E.; {Friedlander}, D.P.; {Hunter}, S.D.;
  {McDonald}, L.M.; et al.~{The Third EGRET Catalog of High-Energy Gamma-Ray Sources}.
\newblock {\em Astrophys. J. Suppl. Ser.} {\bf 1999}, {\em
  123},~79--202,
\newblock
  doi:{\changeurlcolor{black}\href{https://doi.org/10.1086/313231}{\detokenize{10.1086/313231}}}.

\bibitem[{Thompson}(2008)]{Thompson2008}
{Thompson}, D.J.
\newblock {Gamma ray astrophysics: the EGRET results}.
\newblock {\em Rep. Prog. Phys.} {\bf 2008}, {\em 71},~116901,
\newblock
  doi:{\changeurlcolor{black}\href{https://doi.org/10.1088/0034-4885/71/11/116901}{\detokenize{10.1088/0034-4885/71/11/116901}}}.

\bibitem[{Sowards-Emmerd} \em{et~al.}(2003){Sowards-Emmerd}, {Romani}, and
  {Michelson}]{new_id}
{Sowards-Emmerd}, D.; {Romani}, R.W.; {Michelson}, P.F.
\newblock {The Gamma-Ray Blazar Content of the Northern Sky}.
\newblock {\em Astrophys. J.} {\bf 2003}, {\em 590},~109--122,
\newblock
  doi:{\changeurlcolor{black}\href{https://doi.org/10.1086/374981}{\detokenize{10.1086/374981}}}.

\bibitem[{Condon} \em{et~al.}(1998){Condon}, {Cotton}, {Greisen}, {Yin},
  {Perley}, {Taylor}, and~{Broderick}]{nvss}
{Condon}, J.J.; {Cotton}, W.D.; {Greisen}, E.W.; {Yin}, Q.F.; {Perley}, R.A.;
  {Taylor}, G.B.; {Broderick}, J.J.
\newblock {The NRAO VLA Sky Survey}.
\newblock {\em Astron. J.} {\bf 1998}, {\em 115},~1693--1716,
\newblock
  doi:{\changeurlcolor{black}\href{https://doi.org/10.1086/300337}{\detokenize{10.1086/300337}}}.

\bibitem[{Bhattacharya} \em{et~al.}(2003){Bhattacharya}, {Aky{\"u}z}, {Miyagi},
  {Samimi}, and~{Zych}]{EGRET_pos}
{Bhattacharya}, D.; {Aky{\"u}z}, A.; {Miyagi}, T.; {Samimi}, J.; {Zych}, A.
\newblock {On the distribution of EGRET unidentified sources in the Galactic
  plane}.
\newblock {\em Astron. Astrophys.} {\bf 2003}, {\em 404},~163--170,
\newblock
  doi:{\changeurlcolor{black}\href{https://doi.org/10.1051/0004-6361:20030393}{\detokenize{10.1051/0004-6361:20030393}}}.

\bibitem[{Gordon} \em{et~al.}(2016){Gordon}, {Jacobs}, {Beasley}, {Peck},
  {Gaume}, {Charlot}, {Fey}, {Ma}, {Titov}, and~{Boboltz}]{vcs2}
{Gordon}, D.; {Jacobs}, C.; {Beasley}, A.; {Peck}, A.; {Gaume}, R.; {Charlot},
  P.; {Fey}, A.; {Ma}, C.; {Titov}, O.; {Boboltz}, D.
\newblock {Second Epoch VLBA Calibrator Survey Observations: VCS-II}.
\newblock {\em Astron. J.} {\bf 2016}, {\em 151},~154,
\newblock
  doi:{\changeurlcolor{black}\href{https://doi.org/10.3847/0004-6256/151/6/154}{\detokenize{10.3847/0004-6256/151/6/154}}}.

\bibitem[{Charlot} \em{et~al.}(2020){Charlot}, {Jacobs}, {Gordon}, {Lambert},
  {de Witt}, {B\"ohm}, {Fey}, {Heinkelmann}, {Skurikhina}, {Titov}, {Arias},
  {Bolotin}, {Bourda}, {Ma}, {Malkin}, {Nothnagel}, {Mayer}, {MacMillan},
  {Nilsson}, and~R.]{icrf3}
{Charlot}, P.; {Jacobs}, C.S.; {Gordon}, D.; {Lambert}, S.; {de Witt}, A.;
  {B\"ohm}, J.; {Fey}, A.L.; {Heinkelmann}, R.; {Skurikhina},~E.; {Titov}, O.;
  et al.
\newblock {The third realization of the International Celestial Reference Frame
  by very long baseline interferometry}.
\newblock {\em Astron. Astrophys.} {\bf 2020},
\newblock
  doi:{\changeurlcolor{black}\href{https://doi.org/10.1051/0004-6361/202038368}{\detokenize{10.1051/0004-6361/202038368}}}.

\bibitem[{Wright}(2006)]{cosmo-calc}
{Wright}, E.L.
\newblock {A Cosmology Calculator for the World Wide Web}.
\newblock {\em Publ. Astron. Soc. Pac.} {\bf
  2006}, {\em 118},~1711--1715,
\newblock
  doi:{\changeurlcolor{black}\href{https://doi.org/10.1086/510102}{\detokenize{10.1086/510102}}}.

\bibitem[{Beasley} and~{Conway}(1995)]{phase-ref}
{Beasley}, A.J.; {Conway}, J.E.
\newblock {VLBI Phase-Referencing}.
\newblock In \emph{Very Long Baseline Interferometry and~the VLBA}; {Zensus},~J.A.,  {Diamond}, P.J., {Napier}, P.J., Eds.;  {Astronomical Society of the Pacific Conference Series}; Astronomical Society of the Pacific: San Francisco, CA, USA, 1995; Volume~82, p. 327.

\bibitem[{Frey} \em{et~al.}(2016){Frey}, {Paragi}, {Gab{\'a}nyi}, and
  {An}]{ef025}
{Frey}, S.; {Paragi}, Z.; {Gab{\'a}nyi}, K.{\'E}.; {An}, T.
\newblock {Four hot DOGs in the microwave}.
\newblock {\em Mon. Not.  R. Astron. Soc.} {\bf 2016},~{\em 455},~2058--2065,
\newblock
  doi:{\changeurlcolor{black}\href{https://doi.org/10.1093/mnras/stv2399}{\detokenize{10.1093/mnras/stv2399}}}.

\bibitem[{Shepherd} \em{et~al.}(1994){Shepherd}, {Pearson}, and
  {Taylor}]{difmap}
{Shepherd}, M.C.; {Pearson}, T.J.; {Taylor}, G.B.
\newblock {DIFMAP: An interactive program for synthesis imaging.}
\newblock \mbox{{\em Bull. Astron. Soc.}} {\bf 1994}, {\em 26},~987--989.

\bibitem[{H{\"o}gbom}(1974)]{clean}
{H{\"o}gbom}, J.A.
\newblock {Aperture Synthesis with a Non-Regular Distribution of Interferometer
  Baselines}.
\newblock {\em Astron.~Astrophys. Suppl.} {\bf 1974}, {\em
  15},~417--426.

\bibitem[{Diamond}(1995)]{aips_datared}
{Diamond}, P.J.
\newblock {VLBI Data Reduction in Practice}.
\newblock In \emph{Very Long Baseline Interferometry and~the VLBA}; {Zensus},~J.A.,  {Diamond}, P.J., {Napier}, P.J., Eds.;  {Astronomical Society of the Pacific Conference Series}; Astronomical Society of the Pacific: San Francisco, CA, USA, 1995; Volume~82, p. 227.
\bibitem[{Greisen}(2003)]{aips}
{Greisen}, E.W. {AIPS, the~VLA, and~the VLBA}.
\newblock In {\em Information Handling in Astronomy---Historical~Vistas}; \mbox{{Heck}, A., Ed.;} Astrophysics and~Space Science
  Library; Springer: Dordrecht, 2003; Volume~285, p.~109, 
\newblock
  doi:{\changeurlcolor{black}\href{https://doi.org/10.1007/0-306-48080-8_7}{\detokenize{10.1007/0-306-48080-8_7}}}.

\bibitem[{Pearson}(1995)]{modfit}
{Pearson}, T.J.
\newblock {Non-Imaging Data Analysis}.
\newblock In \emph{Very Long Baseline Interferometry and~the VLBA}; {Zensus}, J.A.,  {Diamond}, P.J., {Napier}, P.J., Eds.;  {Astronomical Society of the Pacific Conference Series}; Astronomical Society of the Pacific: San Francisco, CA, USA, 1995; Volume~82, p. 267.

\bibitem[{Kovalev} \em{et~al.}(2005){Kovalev}, {Kellermann}, {Lister}, {Homan},
  {Vermeulen}, {Cohen}, {Ros}, {Kadler}, {Lobanov}, {Zensus}, {Kardashev},
  {Gurvits}, {Aller}, and~{Aller}]{smallest_size}
{Kovalev}, Y.Y.; {Kellermann}, K.I.; {Lister}, M.L.; {Homan}, D.C.;
  {Vermeulen}, R.C.; {Cohen}, M.H.; {Ros}, E.; {Kadler},~M.; {Lobanov}, A.P.;
  {Zensus}, J.A.; et al.
\newblock {Sub-Milliarcsecond Imaging of Quasars and~Active Galactic Nuclei.
  IV. Fine-Scale Structure}.
\newblock {\em Astron. J.} {\bf 2005}, {\em 130},~2473--2505,
\newblock
  doi:{\changeurlcolor{black}\href{https://doi.org/10.1086/497430}{\detokenize{10.1086/497430}}}.

\bibitem[{Karamanavis} \em{et~al.}(2016){Karamanavis}, {Fuhrmann}, {Krichbaum},
  {Angelakis}, {Hodgson}, {Nestoras}, {Myserlis}, {Zensus}, {Sievers}, and
  {Ciprini}]{hibak}
{Karamanavis}, V.; {Fuhrmann}, L.; {Krichbaum}, T.P.; {Angelakis}, E.;
  {Hodgson}, J.; {Nestoras}, I.; {Myserlis}, I.; {Zensus}, J.A.; {Sievers}, A.;
  {Ciprini}, S.
\newblock {PKS 1502+106: A high-redshift Fermi blazar at extreme angular
  resolution. Structural dynamics with VLBI imaging up to 86 GHz}.
\newblock {\em Astron. Astrophys.} {\bf 2016}, {\em 586},~A60,
\newblock
  doi:{\changeurlcolor{black}\href{https://doi.org/10.1051/0004-6361/201527225}{\detokenize{10.1051/0004-6361/201527225}}}.

\bibitem[{Lister} \em{et~al.}(2009){Lister}, {Cohen}, {Homan}, {Kadler},
  {Kellermann}, {Kovalev}, {Ros}, {Savolainen}, and~{Zensus}]{poshiba}
{Lister}, M.L.; {Cohen}, M.H.; {Homan}, D.C.; {Kadler}, M.; {Kellermann}, K.I.;
  {Kovalev}, Y.Y.; {Ros}, E.; {Savolainen},~T.; {Zensus}, J.A.
\newblock {MOJAVE: Monitoring of Jets in Active Galactic Nuclei with VLBA
  Experiments. VI. Kinematics Analysis of a Complete Sample of Blazar Jets}.
\newblock {\em Astron. J.} {\bf 2009}, {\em 138},~1874--1892,
\newblock
  doi:{\changeurlcolor{black}\href{https://doi.org/10.1088/0004-6256/138/6/1874}{\detokenize{10.1088/0004-6256/138/6/1874}}}.

\bibitem[{Fomalont}(1999)]{hiba_kepletek}
{Fomalont}, E.B.
\newblock {Image Analysis}.
\newblock In \emph{Synthesis Imaging in Radio Astronomy II}; {Taylor}, G.B., {Carilli}, C.L., {Perley},~R.A., Eds.; {Astronomical Society of the
  Pacific Conference Series}; Astronomical Society of the Pacific: San Francisco, CA, USA, 1999; Volume 180,  p. 301.

\bibitem[{Lister} \em{et~al.}(2016){Lister}, {Aller}, {Aller}, {Homan},
  {Kellermann}, {Kovalev}, {Pushkarev}, {Richards}, {Ros}, and
  {Savolainen}]{Lister_2016}
{Lister}, M.L.; {Aller}, M.F.; {Aller}, H.D.; {Homan}, D.C.; {Kellermann},
  K.I.; {Kovalev}, Y.Y.; {Pushkarev}, A.B.; {Richards},~J.L.; {Ros}, E.;
  {Savolainen}, T.
\newblock {MOJAVE: XIII. Parsec-scale AGN Jet Kinematics Analysis Based on 19
  years of VLBA Observations at 15 GHz}.
\newblock {\em Astron. J.} {\bf 2016}, {\em 152},~12,
\newblock
  doi:{\changeurlcolor{black}\href{https://doi.org/10.3847/0004-6256/152/1/12}{\detokenize{10.3847/0004-6256/152/1/12}}}.

\bibitem[{Hervet} \em{et~al.}(2016){Hervet}, {Boisson}, and~{Sol}]{Hervet2016}
{Hervet}, O.; {Boisson}, C.; {Sol}, H.
\newblock {An innovative blazar classification based on radio jet kinematics}.
\newblock {\em Astron.~Astrophys.} {\bf 2016}, {\em 592},~A22.
\newblock
  doi:{\changeurlcolor{black}\href{https://doi.org/10.1051/0004-6361/201628117}{\detokenize{10.1051/0004-6361/201628117}}}.

\bibitem[{Lister} \em{et~al.}(2019){Lister}, {Homan}, {Hovatta}, {Kellermann},
  {Kiehlmann}, {Kovalev}, {Max-Moerbeck}, {Pushkarev}, {Readhead}, {Ros}, and
  {Savolainen}]{Lister_2019}
{Lister}, M.L.; {Homan}, D.C.; {Hovatta}, T.; {Kellermann}, K.I.; {Kiehlmann},
  S.; {Kovalev}, Y.Y.; {Max-Moerbeck},~W.; {Pushkarev}, A.B.; {Readhead},
  A.C.S.; {Ros}, E.; et al.
\newblock {MOJAVE. XVII. Jet Kinematics and~Parent Population Properties of
  Relativistically Beamed Radio-loud Blazars}.
\newblock {\em Astrophys. J.} {\bf 2019}, {\em 874},~43,
\newblock
  doi:{\changeurlcolor{black}\href{https://doi.org/10.3847/1538-4357/ab08ee}{\detokenize{10.3847/1538-4357/ab08ee}}}.

\bibitem[{Condon} \em{et~al.}(1982){Condon}, {Condon}, {Gisler}, and
  {Puschell}]{feny.hom}
{Condon}, J.J.; {Condon}, M.A.; {Gisler}, G.; {Puschell}, J.J.
\newblock {Strong radio sources in bright spiral galaxies. II. Rapid star
  formation and~galaxy-galaxy interactions.}
\newblock {\em Astrophys. J.} {\bf 1982}, {\em 252},~102--124,
\newblock
  doi:{\changeurlcolor{black}\href{https://doi.org/10.1086/159538}{\detokenize{10.1086/159538}}}.

\bibitem[{Readhead}(1994)]{equipartition}
{Readhead}, A.C.S.
\newblock {Equipartition Brightness Temperature and~the Inverse Compton
  Catastrophe}.
\newblock {\em Astrophys. J.} {\bf 1994}, {\em 426},~51,
\newblock
  doi:{\changeurlcolor{black}\href{https://doi.org/10.1086/174038}{\detokenize{10.1086/174038}}}.

\bibitem[{Cohen} \em{et~al.}(2007){Cohen}, {Lane}, {Cotton}, {Kassim}, {Lazio},
  {Perley}, {Condon}, and~{Erickson}]{VLSS}
{Cohen}, A.S.; {Lane}, W.M.; {Cotton}, W.D.; {Kassim}, N.E.; {Lazio}, T.J.W.;
  {Perley}, R.A.; {Condon}, J.J.; {Erickson}, W.C.
\newblock {The VLA Low-Frequency Sky Survey}.
\newblock {\em Astron. J.} {\bf 2007}, {\em 134},~1245--1262,
\newblock
  doi:{\changeurlcolor{black}\href{https://doi.org/10.1086/520719}{\detokenize{10.1086/520719}}}.

\bibitem[{Rengelink} \em{et~al.}(1997){Rengelink}, {Tang}, {de Bruyn}, {Miley},
  {Bremer}, {Roettgering}, and~{Bremer}]{WN}
{Rengelink}, R.B.; {Tang}, Y.; {de Bruyn}, A.G.; {Miley}, G.K.; {Bremer}, M.N.;
  {Roettgering}, H.J.A.; {Bremer}, M.A.R.
\newblock {The Westerbork Northern Sky Survey (WENSS), I. A 570 square degree
  Mini-Survey around the North Ecliptic Pole}.
\newblock {\em Astron. Astrophys. Suppl.} {\bf 1997}, {\em
  124},~259--280,
\newblock
  doi:{\changeurlcolor{black}\href{https://doi.org/10.1051/aas:1997358}{\detokenize{10.1051/aas:1997358}}}.

\bibitem[{Douglas} \em{et~al.}(1996){Douglas}, {Bash}, {Bozyan}, {Torrence},
  and~{Wolfe}]{TXS}
{Douglas}, J.N.; {Bash}, F.N.; {Bozyan}, F.A.; {Torrence}, G.W.; {Wolfe}, C.
\newblock {The Texas Survey of Radio Sources Covering $-35.5$ degrees <
  declination < 71.5 degrees at 365 MHz}.
\newblock {\em Astron. J.} {\bf 1996}, {\em 111},~1945,
\newblock
  doi:{\changeurlcolor{black}\href{https://doi.org/10.1086/117932}{\detokenize{10.1086/117932}}}.

\bibitem[{Colla} \em{et~al.}(1973){Colla}, {Fanti}, {Fanti}, {Ficarra},
  {Formiggini}, {Gandolfi}, {Gioia}, {Lari}, {Marano}, {Padrielli}, and
  {Tomasi}]{B2.3}
{Colla}, G.; {Fanti}, C.; {Fanti}, R.; {Ficarra}, A.; {Formiggini}, L.;
  {Gandolfi}, E.; {Gioia}, I.; {Lari}, C.; {Marano}, B.; {Padrielli},~L.; et al.
\newblock {The B2 catalogue of radio sources - third part}.
\newblock {\em Astron. Astrophys. Suppl.} {\bf 1973}, {\em 11},~291.

\bibitem[{Gregory} and~{Condon}(1991)]{87GB}
{Gregory}, P.C.; {Condon}, J.J.
\newblock {The 87GB Catalog of Radio Sources Covering 0 degrees < delta < +75
  degrees at 4.85 GHz}.
\newblock {\em Astrophys. J. Suppl. Ser.} {\bf 1991}, {\em
  75},~1011,
\newblock
  doi:{\changeurlcolor{black}\href{https://doi.org/10.1086/191559}{\detokenize{10.1086/191559}}}.

\bibitem[{Myers} \em{et~al.}(2003){Myers}, {Jackson}, {Browne}, {de Bruyn},
  {Pearson}, {Readhead}, {Wilkinson}, {Biggs}, {Blandford}, {Fassnacht},
  {Koopmans}, {Marlow}, {McKean}, {Norbury}, {Phillips}, {Rusin}, {Shepherd},
  and~{Sykes}]{CLASS}
{Myers}, S.T.; {Jackson}, N.J.; {Browne}, I.W.A.; {de Bruyn}, A.G.; {Pearson},
  T.J.; {Readhead}, A.C.S.; {Wilkinson}, P.N.; {Biggs}, A.D.; {Blandford},
  R.D.; {Fassnacht}, C.D.; et al.
\newblock {The Cosmic Lens All-Sky Survey - I. Source selection and
  observations}.
\newblock {\em Mon. Not.  R. Astron. Soc.} {\bf 2003},
  {\em 341},~1--12,
\newblock
  doi:{\changeurlcolor{black}\href{https://doi.org/10.1046/j.1365-8711.2003.06256.x}{\detokenize{10.1046/j.1365-8711.2003.06256.x}}}.

\bibitem[{Kim} and~{Trippe}(2014)]{vimap}
{Kim}, J.Y.; {Trippe}, S.
\newblock {VIMAP: An Interactive Program Providing Radio Spectral Index Maps of
  Active Galactic Nuclei}.
\newblock {\em J. Korean Astron. Soc.} {\bf 2014}, {\em
  47},~195--199,
\newblock
  doi:{\changeurlcolor{black}\href{https://doi.org/10.5303/JKAS.2014.47.5.195}{\detokenize{10.5303/JKAS.2014.47.5.195}}}.

\bibitem[{Casandjian} and~{Grenier}(2008)]{alternate_3EG}
{Casandjian}, J.M.; {Grenier}, I.A.
\newblock {A revised catalogue of EGRET {\ensuremath{\gamma}}-ray sources}.
\newblock {\em Astron. Astrophys.} {\bf 2008},~{\em 489},~849--883,
\newblock
  doi:{\changeurlcolor{black}\href{https://doi.org/10.1051/0004-6361:200809685}{\detokenize{10.1051/0004-6361:200809685}}}.

\bibitem[{Bhattacharya} \em{et~al.}(2017){Bhattacharya}, {Mohana A}, {Gulati},
  {Bhattacharyya}, {Bhatt}, {Sreekumar}, and~{Stalin}]{egret_fermi_var_2017}
{Bhattacharya}, D.; {Mohana A}, K.; {Gulati}, S.; {Bhattacharyya}, S.; {Bhatt},
  N.; {Sreekumar}, P.; {Stalin}, C.S.
\newblock {Unusual long-term low-activity states of EGRET blazars in the Fermi
  era}.
\newblock {\em Mon. Not.  R. Astron. Soc.} {\bf 2017},
  {\em 471},~5008--5017,
\newblock
  doi:{\changeurlcolor{black}\href{https://doi.org/10.1093/mnras/stx1827}{\detokenize{10.1093/mnras/stx1827}}}.

\end{thebibliography}



\end{document}